\documentclass[12pt]{iopart} 

\usepackage{graphicx}
\usepackage{rotating}
\usepackage{amssymb}

\newcommand{\la}{\mathord{\langle}}
\newcommand{\ra}{\mathord{\rangle}}

\begin{document}

\title
[Low-energy properties of two-dimensional magnetic nanostructures] 
{Low-energy properties of two-dimensional magnetic nanostructures: 
interparticle interactions and disorder effects} 

\author
{P J Jensen\footnote{Corresponding author. On leave from: Institut f\"ur 
Theoretische Physik, Freie Universit\"at Berlin, Arnimallee 14, D-14195 
Berlin, Germany.} and G M Pastor} 

\address{Laboratoire de Physique Quantique, Universit\'e Paul Sabatier, \\
Centre National de la Recherche Scientifique, \\ 
118, route de Narbonne, F-31062 Toulouse, France } 

\eads{\mailto{jensen@physik.fu-berlin.de} \\  
\mailto{gustavo.pastor@irsamc.ups-tlse.fr}} \vspace{0.5cm}  

\hspace*{1.7cm}{\small \today}\\ 

\date
{\today}

\begin{abstract} 
The low-energy properties of 
two-dimensional ensembles of dipole-coupled magnetic nanoparticles
are studied as function of structural disorder and particle coverage. 
Already small deviations from a square particle arrangement 
lift the degeneracies of the microvortex magnetic configuration, 
and result in a strongly noncollinear 
magnetic order of the particle ensemble. 
The energy distribution of metastable states is determined. 
For a low degree of disorder a strongly asymmetric shape with a pronounced 
peak of the ground state energy results. In contrast, 
for a strong disorder a Gaussian-like distribution is obtained. 
The average dipole coupling energy $\overline E_\mathrm{dip}$  
decreases with increasing structural disorder.  
The role of vacancies has been studied for a square particle array 
by determining the angular distribution of the preferred microvortex 
angle as function of the vacancy concentration. 
Indications for a preferred angular direction along the axial as well
as along the diagonal directions of the square array are revealed. 
A corresponding investigation for disturbed square arrays results 
in a different angular distribution. 
The effect of dipole-quadrupole corrections resulting from the finite size 
of the particles is quantified. 
\end{abstract}


\submitto{\NJP}

\maketitle

\section{Introduction} \label{sec:intro}
Interacting magnetic nanostructured materials 
are currently the subject of intense research activity, 
driven by their fundamental interest and technological 
perspectives \cite{Dor97}. Numerous experimental and theoretical studies
have been performed for various two-dimensional (2D) \cite{Sun00} and 
three-dimensional (3D) \cite{Luo91,Zha96,And97,Han98,Dor99,Lom00} 
arrangements of nanometer-size magnetic particles having
different degrees of structural and magnetic disorder.
The magnetic behaviour of these systems is determined by 
single-particle properties (e.g., particle moments, lattice and shape 
anisotropies, etc.), by the composition and morphology of the nanostructure,
and in particular by the nature of the dominant interparticle interactions. 
The latter comprises especially the magnetic dipole coupling, which 
will be adressed in the present study. Other interparticle
interactions are, for example, the Rudermann-Kittel-Kasuya-Yosida (RKKY) 
indirect exchange mediated by the 
conduction electrons of a metallic substrate, or the 
short-range direct exchange in case when the particles are in contact. 
The relative importance of single-particle versus interparticle contributions 
can be tuned experimentally at least to some extent 
by changing sample characteristics such as the particle-size 
distribution or the average interparticle distance. 
For low particle coverages 
the interactions can be treated as a perturbation to the
single-particle properties. However, for dense particle ensembles 
the interparticle interactions become increasingly important and 
eventually dominate. In this interesting case the single-particle
approach is no longer applicable  and an explicit treatment of the 
interactions is unavoidable \cite{Luo91,Zha96,And97}. 

A fundamental question in this context is to identify and understand the 
collectively ordered magnetic states which are induced by the 
interactions in such particle ensembles. A few basic properties 
of strictly periodic dipole-coupled systems are summarized. 
The ground state of a \textit{square} lattice of equal-sized 
(monodispersed) particles 
is the so-called \textit{microvortex} (MV) magnetic 
arrangement \cite{BIG83,PrH90}. This magnetic structure is 
characterized by the microvortex angle $\phi_\mathrm{mv}$, 
where the angles of the particle magnetizations of 
a plaquette of four neighbouring particles are given by 
$\phi_1=\phi_\mathrm{mv}$, $\phi_2=-\phi_\mathrm{mv}$, 
$\phi_3=180^\circ+\phi_\mathrm{mv}$, and 
$\phi_4=180^\circ-\phi_\mathrm{mv}$. In particular, multiples of
$\phi_\mathrm{mv}=90^\circ$ represent 
\textit{columnar} states, consisting of ferromagnetic 
rows or columns with alternating signs of magnetizations. 
Evidently, the MV state has a vanishing net magnetization.  
The parallel or ferromagnetic (FM) state has a larger energy and 
is actually an unstable solution. The ordering of dipole-coupled 
spins is very sensitive to the lattice structure. For example, 
the honeycomb lattice has a ground state with a vanishing net 
magnetization which is similar to the MV state of the square
lattice \cite{PrH90}. In contrast, for the hexagonal lattice 
the ground state is ferromagnetic \cite{ROC91,Pol02}. 

Despite the fact that the dipole interaction is not  rotationally invariant, 
the ground state of these dipole-coupled periodic particle arrays are 
continuously degenerate with respect to a rotation of the MV angle 
$\phi_\mathrm{mv}$ or of the FM angle $\phi_\mathrm{fm}$, respectively. 
This holds for classical spins at $T=0$. 
Thermal fluctuations, quantum fluctuations, 
or a structural disorder immediately lift  
these accidental ground state degeneracies of the periodic 
structures. The energy lowering associated to the symmetry breaking
stabilizes the system in some particular magnetic order. Therefore, 
the square and honeycomb lattices are said to show 
a so-called 'order-by-disorder' effect \cite{PrH90,Vil79}. 
This implies that internal (nontrivial) degeneracies (e.g., 
the relative directions of magnetic sublattices) are lifted by the
presence of disorder, and that particular directions of the sublattice   
magnetizations are preferred. Moreover, for a dipole-coupled square lattice 
the fluctuations induce a 
fourfold in-plane magnetic anisotropy. As shown by Prakash 
and Henley \cite{PrH90}, the preferred in-plane magnetic orientations  
are the axial directions (columnar states) 
for thermal and quantum fluctuations, and the 
diagonal directions for a structural disorder induced by a small amount 
of vacancies. In addition, Monte  Carlo calculations and 
interacting spin  wave theory indicate that a magnetic ordering at finite  
temperatures exists for a dipole-coupled square spin lattice, 
since  the magnetic excitations are not continuously 
degenerate \cite{DMB97}. Thus, the Mermin-Wagner theorem, which 
excludes an ordered state 
for a continuously degenerate 2D system at finite temperatures \cite{MeW66}, 
is not effective in this case. Notice that already the 
\textit{long-range} character of the dipole interaction induces a 
collectively ordered magnetic state in a square lattice \cite{YKG86}. 

Structural disorder, which is usually present in real magnetic nanostructures,
results for example from the size and shape dispersion, positional 
disorder, or random anisotropy axes. Due to the nonuniform and 
competing nature of the magnetic couplings the magnetic ordering 
in disordered particle ensembles is similar to 
the one of a spin-glass \cite{Bin86}. Thus, many different metastable 
states exist, which are characterized by strong magnetic noncollinearities. 
An intriguing question in this context is whether the dipole interaction 
in \textit{disordered} planar particle ensembles results in a 
collectively ordered magnetic state. 
Several experiments on interacting, high-density ferrofluid systems indicate 
the onset of a collectively ordered state below a characteristic, 
concentration-dependent temperature. For example, at this temperature 
a 'critical slowing down' of the magnetic relaxation is observed 
\cite{Luo91,Zha96}. Furthermore, recent 
measurements on Co islands on Cu(001) exhibit 
a magnetic hysteresis and remanence in the temperature range up to 150 K 
also for coverages below the magnetic percolation 
threshold \cite{Bov99}. Due to the small size of the Co islands these 
findings could not explained simply by single-particle blocking effects. 
Note that the experimental determination of the ordering temperature 
is difficult, since the relaxation times are often very long. 
Hence, it is of considerable interest to analyze how 
the magnetic order depends on the sample parameters which can be
controlled in experiment. 

The purpose of this paper is the theoretical study of the 
low-energy properties of dipole-coupled magnetic particle ensembles 
forming inhomogeneous planar arrangements with various degrees of 
disorder ranging from a quasi-periodic square 
lattice to a random array. Numerical simulations are performed 
in order to achieve a detailed microscopic description within the
model. The low symmetry of the system and the complicated nature of the
interaction seem to preclude simple analytical approaches. 
We focus on the strongly interacting case, neglecting 
single-particle anisotropies. 
Special attention is paid to the role of disorder and 
noncollinear arrangements of the particle magnetic moments. 
In particular, 
we determine the energy distribution of metastable states and the 
average magnetic dipole energy for different coverages and types of 
disorder. Furthermore, we investigate global and local 
order parameters as function of disorder, and compare our results
with previous calculations \cite{PrH90}. 
All calculations are performed at $T=0$. 

The rest of the paper is organized as follows.
The theoretical methods are outlined in section \ref{sec:theo}. 
Representative results for the magnetic arrangements, as well as for  
various magnetic properties are presented in section \ref{sec:res}. 
A conclusion is given in section \ref{sec:conc}. Finally, 
in the Appendix we report details for 
the extension of the dipole summation 
beyond the point-dipole approximation which takes 
into account effects resulting from the finite particle size.  

\section{Theory} \label{sec:theo}
We consider a 2D rectangular unit cell in the $xy$- plane with 
$n=n_x\times n_y$ non-overlapping, disk-shaped magnetic particles. 
Due to the strong direct exchange interaction and the small size of
the particles under consideration, each particle $i$ can be viewed 
as a single magnetic domain (Stoner-Wohlfarth particle) \cite{StW48}. 
Thus, a particle containing $N_i$ atoms  
carries a giant spin $M_i = N_i\,\mu_{at}$, where $\mu_{at}$ is 
the atomic magnetic moment. For simplicity, we restrict the particle 
magnetizations to be confined to the $xy$ plane, 
the planar rotator $\bi{M}_i$ is then characterized by the 
in-plane angle $\phi_i$: 
$\bi{M}_i=(M_i^x,M_i^y,M_i^z)=M_i(\cos\phi_i,\sin\phi_i,0)$. 
In this study no size dispersion is considered, i.e., $N_i=N$. 
For disk-shaped particles
the particle radius $r_0$ is given by $r_0/a_0=\sqrt{N}$, where $a_0$ 
is the interatomic distance. Unless otherwise stated, the present 
results refer to a unit cell containing $n=100$ particles with
$N=1000$ atoms each. The size of the unit cell is given by 
$L_x\times L_y=(n_x\,R_0)\times(n_y\,R_0)$, with $R_0$ 
the average interparticle distance. 
For a planar array of circular particles the overall surface 
coverage is $C=\pi(r_0/R_0)^2$. 
Four different types of the lateral particle arrangement 
have been considered: (i) a periodic square array, i.e., the particle 
centers are located on the sites 
of a square lattice with lattice constant $R_0$, (ii) a disturbed 
(quasi-periodic) array for which the particle centers deviate 
randomly from the square array, using a Gaussian distribution 
$P(\bi{R})$ with positional standard deviation $\sigma_R$, 
(iii) a diluted square particle lattice containing a number of vacancies
$n_\mathrm{vac}$ with concentration $C_\mathrm{vac}$, 
and (iv) a fully random distribution of non-overlapping particles 
within the unit cell. 
Periodic boundary conditions are introduced in order to describe 
an infinitely extended planar particle ensemble. 
By considering a single unit cell a finite  system
can be realized as well. In this case the
effect of boundaries will strongly dominate the
resulting magnetic arrangement. 

For such particle ensembles we consider the dipole-dipole 
interaction between the magnetic moments $\bi{M}_i$, which can 
also be expressed in terms of a dipole field $\bi{B}_i^\mathrm{dip}$ 
acting on $\bi{M}_i$ due to all other particle magnetic moments: 
\begin{equation}
E_\mathrm{dip}=\frac{\mu_0}{2}\;\sum_{i,j \atop i\ne j} 
\left[\bi{M}_i\;\bi{M}_j\;r_{ij}^{-3}-3\big(\bi{r}_{ij}\,\bi{M}_i 
\big)\,\big(\bi{r}_{ij}\,\bi{M}_j\big)\; r_{ij}^{-5}\right] 
=- \frac{\mu_0}{2}\;\sum_i\; \bi{M}_i\;\bi{B}_i^\mathrm{dip} 
\;, \label{e1} \end{equation} 
where $r_{ij}=|\mathbf{r}_{ij}|=|\mathbf{r}_i-\mathbf{r}_j|$ is 
the distance between the centers of particles $i$ and $j$, and 
$\mu_0$ the vacuum permeability.  
The infinite range of the dipole interaction is taken into account 
by applying an Ewald-type summation over all periodically arranged 
unit cells of the extended thin film \cite{Jen97}. 
In addition to the usual point-dipole sum we consider the leading 
correction resulting from the finite particle size 
(dipole-quadrupole interaction), which is outlined in detail in the 
Appendix. This correction becomes comparable to the 
point-dipole sum for large particle coverages or small interparticle 
distances. The energy unit of the dipole coupling is given by 
$E_\mathrm{dip}^0=\mu_{at}^2/a_0^3$, with $\mu_{at}$ in units of 
the Bohr magneton $\mu_B$. In this study we assume values appropriate 
to Fe ($\mu_{at}=2.2\;\mu_B$, $a_0=2.5$~\AA), which yields  
$E_\mathrm{dip}^0=0.19$~K. The effects of 
single-particle anisotropies resulting from the spin-orbit interaction, 
the dipole interaction among atomic magnetic moments within each 
particle (shape anisotropy), external magnetic fields, and 
finite temperatures are beyond the scope of the present study. 

Starting from an arbitrary initial configuration $\{\phi_i^\mathrm{initial}\}$ 
of the magnetic directions, the total magnetic energy $E_\mathrm{dip}$
of the system is relaxed to the nearest local minimum, which often  
corresponds to a metastable state, by varying all in-plane angles 
$\phi_i$ of the 
particles using a conjugated gradient method \cite{CGM}. 
For example, the experimental situation of a remanent state after 
removal of an external magnetic field is simulated by choosing 
a fully aligned initial arrangement along a certain direction.
We emphasize that the applied procedure is not intended to 
search preferently for the global energy minimum or ground state, 
which is the equilibrium state at $T=0$. Rather, at first we 
determine the energy distribution of the local minima 
for different degrees of structural disorder introduced in the planar 
particle array. A number of randomly chosen intital 
configurations $\{\phi_i^\mathrm{initial}\}$ is created and relaxed to a 
nearby local minimum. This metastable 
state is characterized by its energy $E_\mathrm{dip}$ and by its set 
of angles \{$\phi_i$\}. A twofold (uniaxial) symmetry 
is always present due to time inversion symmetry, i.e., the energy of 
a state does not change under the 
transformation $\phi_i\to\phi_i+180^\circ$ performed simultaneously 
for all particles. These mirrored states are considered to be
equivalent. 
The numbers of trials yielding relaxed states with energies falling 
into given energy intervals are monitored, and the corresponding 
energy histogram is determined. 
Since the minimization procedure relaxes typically to the
local energy minimum that is closest to the initial state, 
this frequency provides a measure of the 
\textit{catchment area,} i.e., the area of the 'energy valley'
belonging to that state in the high-dimensional configuration space. 

In order to account for the large number of local energy minima 
occurring in particle arrangements with structural inhomogeneities 
we determine the \textit{average dipole energy} 
$\overline E_\mathrm{dip}$ resulting from many different trials
for the same particle array. In addition, 
we average $\overline E_\mathrm{dip}$ over a number of different 
realizations of the unit 
cell, using the same global variables which characterize the particle 
ensemble (coverage, standard deviation $\sigma_R$, etc.). 
For comparison, $\overline E_\mathrm{dip}$ is also calculated for the 
same spatial setup by assuming a ferromagnetic state 
(i.e., $\phi_i=\phi_\mathrm{fm}$) and averaging over the FM
angle $\phi_\mathrm{fm}$. In case of a disturbed (quasi-periodic) square 
lattice, i.e., if the 
particle numbering allows for the identification of a square plaquette 
of four neighbouring particles, we also determine
the dipole energy for the microvortex magnetic arrangement, 
averaging over the MV angle $\phi_\mathrm{mv}$. Note that for a 
nonuniform particle ensemble these FM and MV  
magnetic arrangements do not correspond in general to local energy minima. 
For a \textit{random} set of angles \{$\phi_i$\} one obtains 
$\overline E_\mathrm{dip}=0$ which constitutes the natural energy 
reference. 

The deviations of the metastable low-energy magnetic arrangements 
from the MV state are quantified by the \textit{global} and 
\textit{local microvortex order parameters} 
$M_\mathrm{mv}^\mathrm{global}(\sigma_R)$ and 
$M_\mathrm{mv}^\mathrm{local}(\sigma_R)$ given by 
\begin{eqnarray}  
M_\mathrm{mv}^\mathrm{global}(\sigma_R) &=& \frac{1}{n}\bigg[\bigg(
\sum_i^n(-1)^{i_y}\cos\phi_i\bigg)^2+\bigg(\sum_i^n
(-1)^{i_x}\sin\phi_i\bigg)^2\bigg]^{1/2} \;, \label{e1a} \\ 
M_\mathrm{mv}^\mathrm{local}(\sigma_R) &=& \frac{1}{4}\bigg[ \bigg( 
\sum_i^4(-1)^{i_y}\cos\phi_i\bigg)^2+\bigg(\sum_i^4
(-1)^{i_x}\sin\phi_i\bigg)^2 \bigg]^{1/2} \;, \label{e1b} \end{eqnarray} 
where $i_y$ and $i_x$ denote the numbering of the rows and columns  
corresponding to particle $i$. Clearly, a reference square lattice 
is prerequisitive. Hence, we restrict ourselves to a quasi-periodic 
array or to a square array with vacancies for which   $i_y$ and $i_x$ 
can be uniquely defined. These two order parameters 
differ by the sum $i$ running either over all $n$ particles of the 
unit cell or over the four neighbouring particles within a square 
plaquette. They have a simple geometrical interpretation as the 
projection of the magnetic configuration on 
the two linear independent columnar states having 
$\phi_\mathrm{mv}=0^\circ$ and $\phi_\mathrm{mv}=90^\circ$,  
regardless of all possible rotations of the MV states within this plane.
$M_\mathrm{mv}^\mathrm{global}(\sigma_R)$ measures the MV 
order in the nanostructure, whereas 
$M_\mathrm{mv}^\mathrm{local}(\sigma_R)$ measures the short range 
ordering. Both quantities are averaged over an appropriate  
number of initial configurations for the same particle 
arrangement and over different realizations of the unit cell. In addition 
$M_\mathrm{mv}^\mathrm{local}(\sigma_R)$ is averaged over all 
four-particle plaquettes within the unit cell. 

Finally, another quantity characterizing the magnetic properties of 
nanostructured particle arrangements is 
the distribution of the microvortex 
angles $\phi_\mathrm{mv}$. Prakash and Henley observed that the 
preferred MV angles for a square particle array with a small number of
randomly distributed vacancies are the diagonal directions 
($\phi_\mathrm{mv}=45^\circ,\;135^\circ$, etc.) \cite{PrH90}.  
It is therefore interesting to investigate how the angular
distribution of the microvortex angles 
depends on the type and degree of disorder. For this
purpose different realizations of the nanostructure are created. For
each particle setup the MV state with angle $\phi_\mathrm{mv}$
yielding the lowest energy is determined, and the corresponding 
frequencies for all angles are monitored. We consider a \textit{disturbed} 
square particle array with standard deviation $\sigma_R$, 
and a \textit{diluted} square particle lattice with a concentration 
$C_\mathrm{vac}$ of vacancies. It should be noted that for a 
small number of vacancies one obtains an appreciable dependence 
on the shape of the unit cell (aspect ratio 
of the rectangle), since these vacancies are located effectively on a 
square or rectangular lattice due to the periodic boundary conditions. 
\begin{figure}[p]
\hspace*{2cm} \includegraphics[bb=30 25 560 810,width=13cm]
        {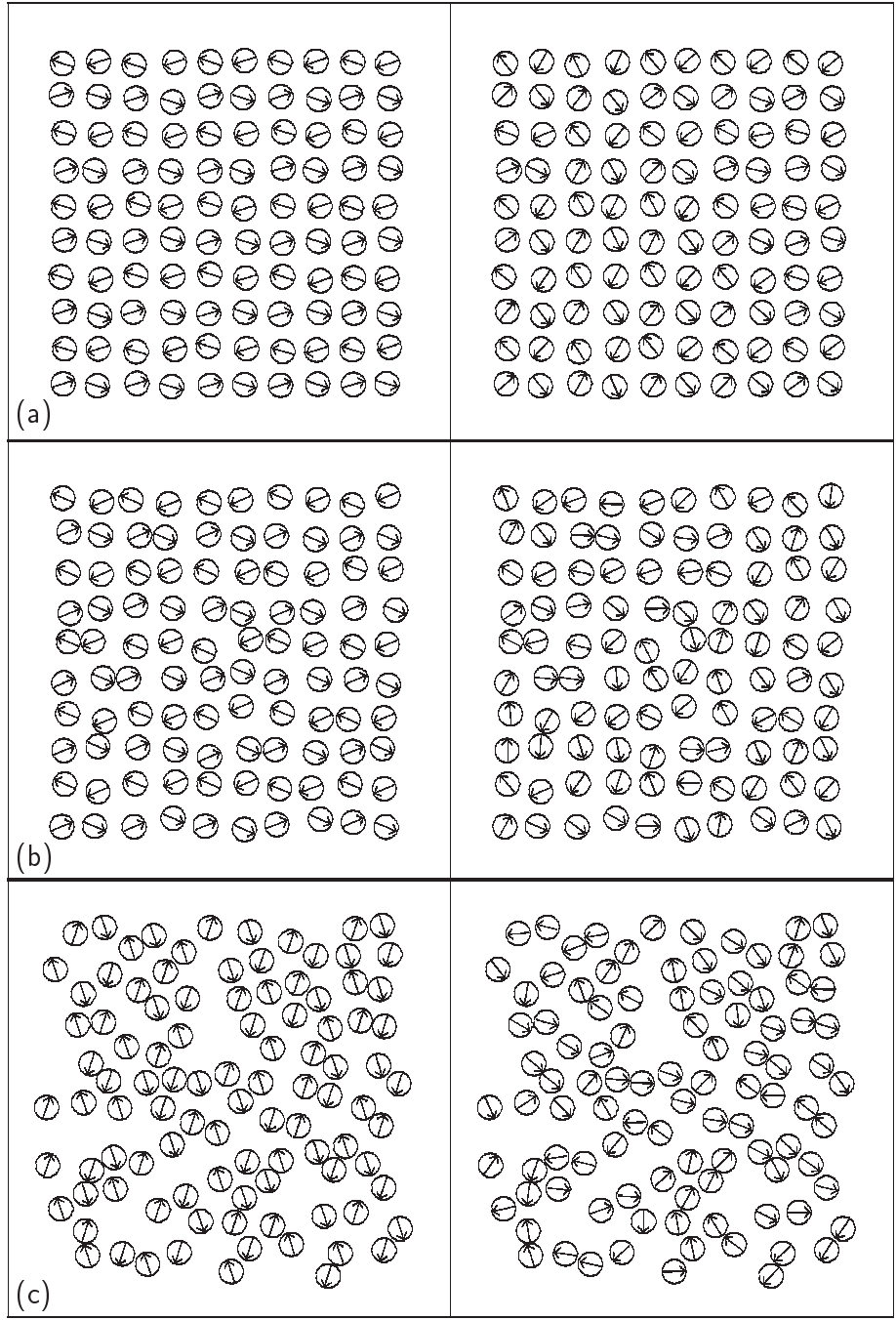}
\caption{
Illustrations of the magnetic arrangements in two-dimensional 
nanostructures. The particle positions are scattered around 
the lattice sites of a square array with positional standard deviations 
(a) $\sigma_R/R_0=0.05$, (b) $\sigma_R/R_0=0.1$, and (c) $\sigma_R/R_0=0.5$,
where $R_0$ refers to the average interparticle distance.  
The surface coverage amounts to $C=35$~\%. The left columns refers to 
microvortex magnetic arrangements. Relaxation of the magnetic moment 
directions using these arrangements as starting
configurations yields the metastable solutions shown in the right
columns. }\end{figure} 

\section{Results} \label{sec:res}
First, figure~1 illustrates some representative low-energy magnetic 
arrangements of disordered particle arrays. 
With increasing disorder, characterized by the positional 
standard deviation $\sigma_R$, the magnetic configurations of the
relaxed solutions become increasingly noncollinear. 
Results are given for the magnetic states of a slightly 
disturbed ($\sigma_R/R_0=0.05$), a moderately disturbed ($\sigma_R/R_0=0.10$), 
and a strongly disturbed ($\sigma_R/R_0=0.50$) particle setup having all the 
same coverage $C=35$~\%. Both the optimal microvortex states 
and the relaxed solutions are shown. 
For $\sigma_R/R_0=0.05$ the MV arrangement resembles quite closely 
the true solution, see figure~1(a). However, as can be seen in 
figure~1(b), already a moderate amount of positional disorder destroys 
the MV state. This is physically reasonable since the MV order is tightly 
connected to the presence of a square-lattice symmetry of the 
particle ensemble. The degree of the MV ordering will be quantified 
below. For large $\sigma_R$ or for a random 
particle setup the resulting magnetic arrangement is dominated by 
the formation of chains and loops of magnetic moments with a correlated  
`head-to-tail' alignment of the particle magnetizations, see figure~1(c)  
\cite{Cha96}. This reflects the tendency of the dipole interaction 
to favour a locally demagnetized state with a vanishing or small 
net magnetization. 
\begin{figure}[t] 
\hspace*{3cm}\includegraphics[bb=40 30 570 765,angle=-90,clip,width=12cm]
        {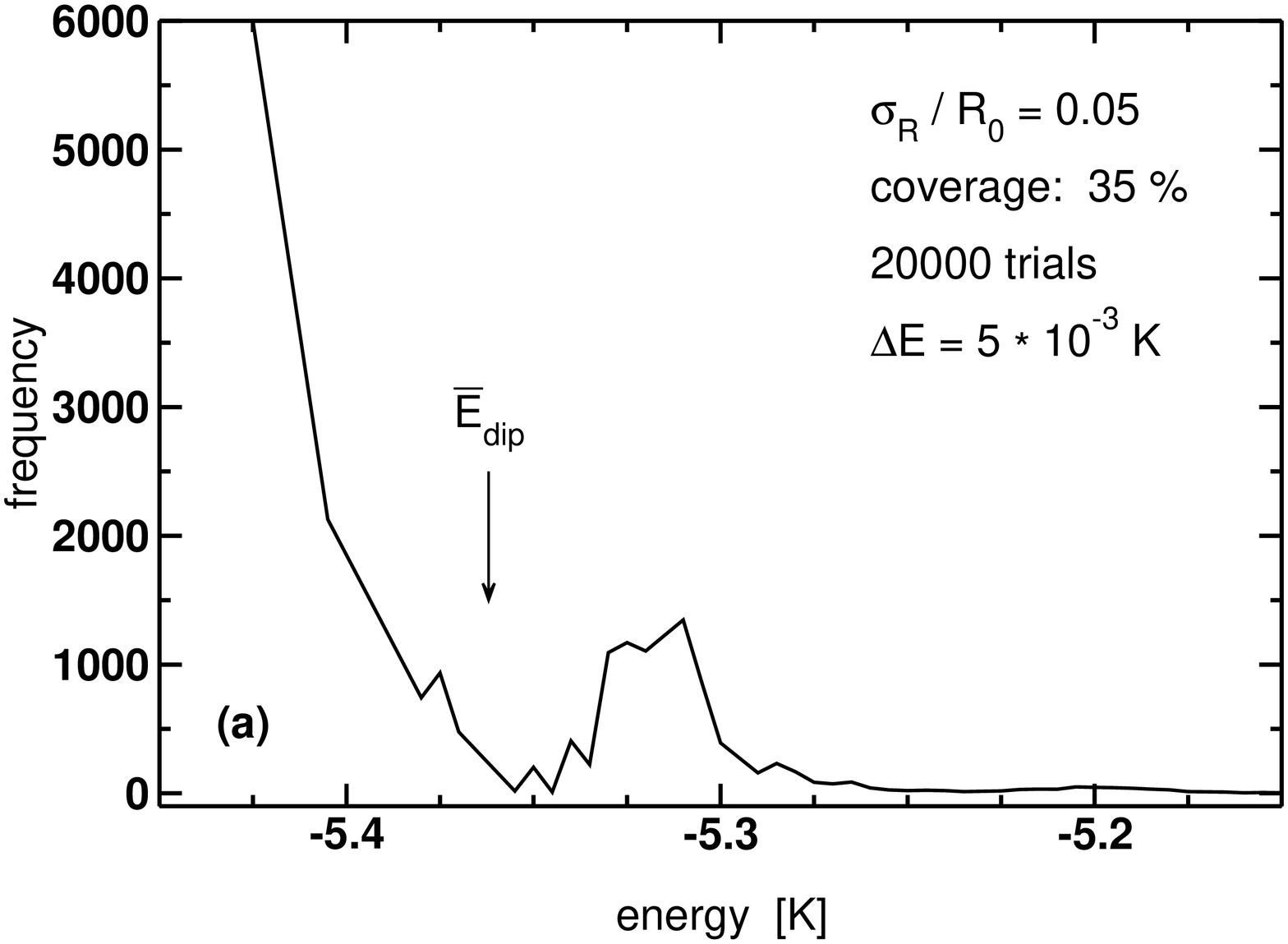}\\
\hspace*{3cm}\includegraphics[bb=40 30 570 765,angle=-90,clip,width=12cm]
        {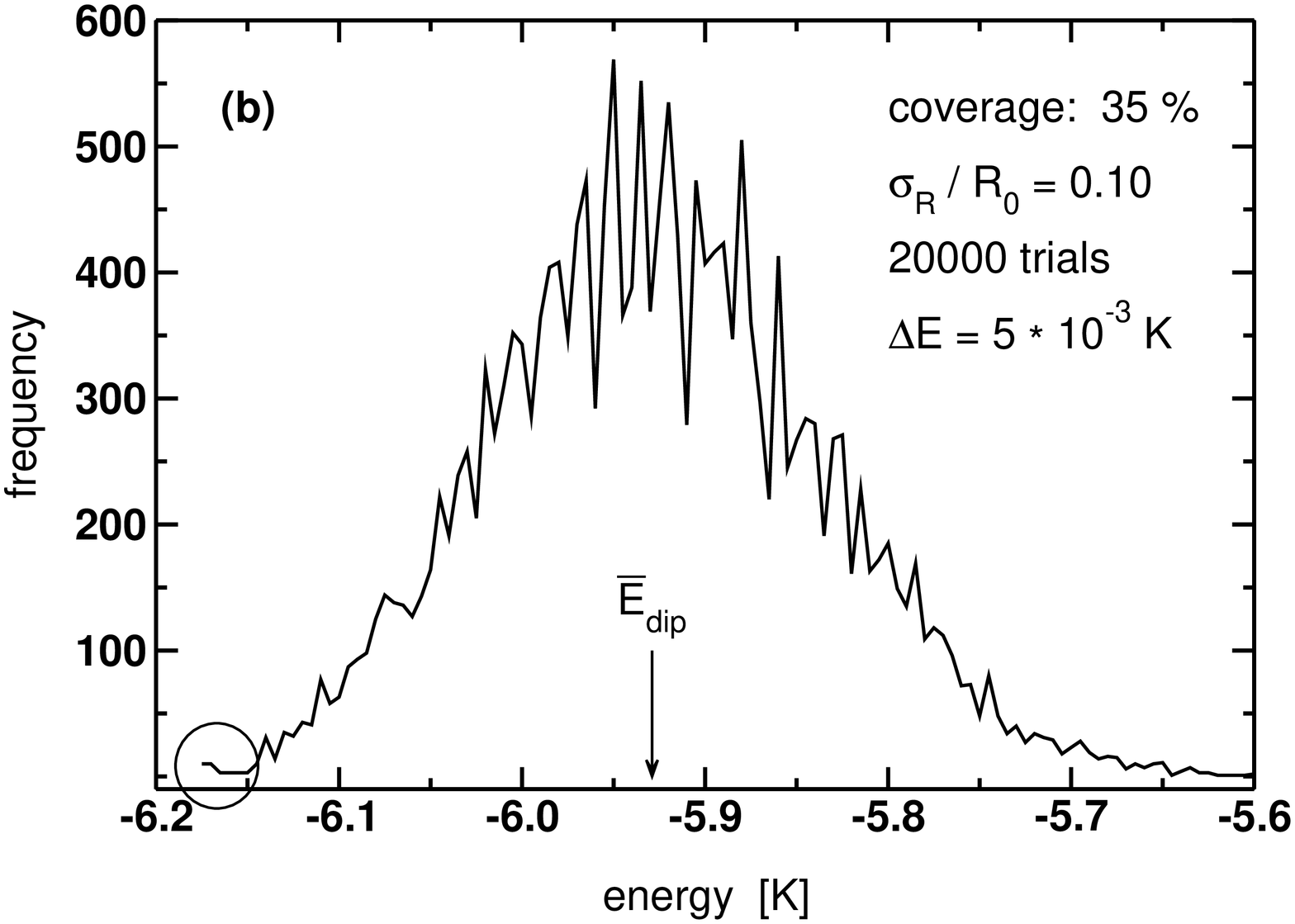}
\caption{
Energy distribution of local energy minima for two disturbed square
particle arrays using different positional standard deviations (a) 
$\sigma_R/R_0=0.05$, and (b) $\sigma_R/R_0=0.10$. The particle coverage
amounts to $C=35$~\%. The distributions of the metastable states are 
obtained from 20000 
randomly chosen initial configurations and sampled into energy intervals 
with width $\Delta E=5\cdot10^{-3}$~K. Also indicated is the resulting 
average dipole energy $\overline E_\mathrm{dip}(\sigma_R)$. 
}\end{figure} 

In figure~2 we present examples for the dipole energy distribution of 
metastable states of slightly and moderately disturbed particle setups 
shown in figures~1(a) and (b). These were obtained by considering 
20000 random initial configurations $\{\phi_i^\mathrm{initial}\}$ 
and by assigning the energies of the relaxed states to energy 
intervals $[E,E+\Delta E]$, with $\Delta E=5\cdot10^{-3}$~K. 
Distinctly different energy distributions are observed for weak and
strong disorder. For weak disorder [e.g., $\sigma_R/R_0=0.05$, 
see figure~2(a)] an \textit{asymmetric} energy
distribution is found. The lowest energy state is 
reached very often, in fact about $50$~\% of the trials relax to 
that ground state for $\sigma_R/R_0=0.03$, and about 
$30$~\% for $\sigma_R/R_0=0.05$. Moreover, 
the ground state configuration resembles closely the MV 
state with an almost vanishing net magnetization. In addition, 
numerous metastable states are obtained, which energies are
distributed over a relatively broad range, and which are reached 
far less frequently. In other words, for weak disturbances the
catchment area of the ground state is much 
larger than the ones of the higher-energy states. 
It is interesting to note that the energy 
differences between the few states with the lowest energies are 
quite larger than those found for less stable magnetic arrangements. 
Since these latter states often exhibit a finite net magnetization, 
one expects that an external magnetic field tends to stabilize these 
higher-energy metastable configurations. 
\begin{figure}[t]
\includegraphics[bb=60 1 600 740,angle=-90,clip,width=9cm]
        {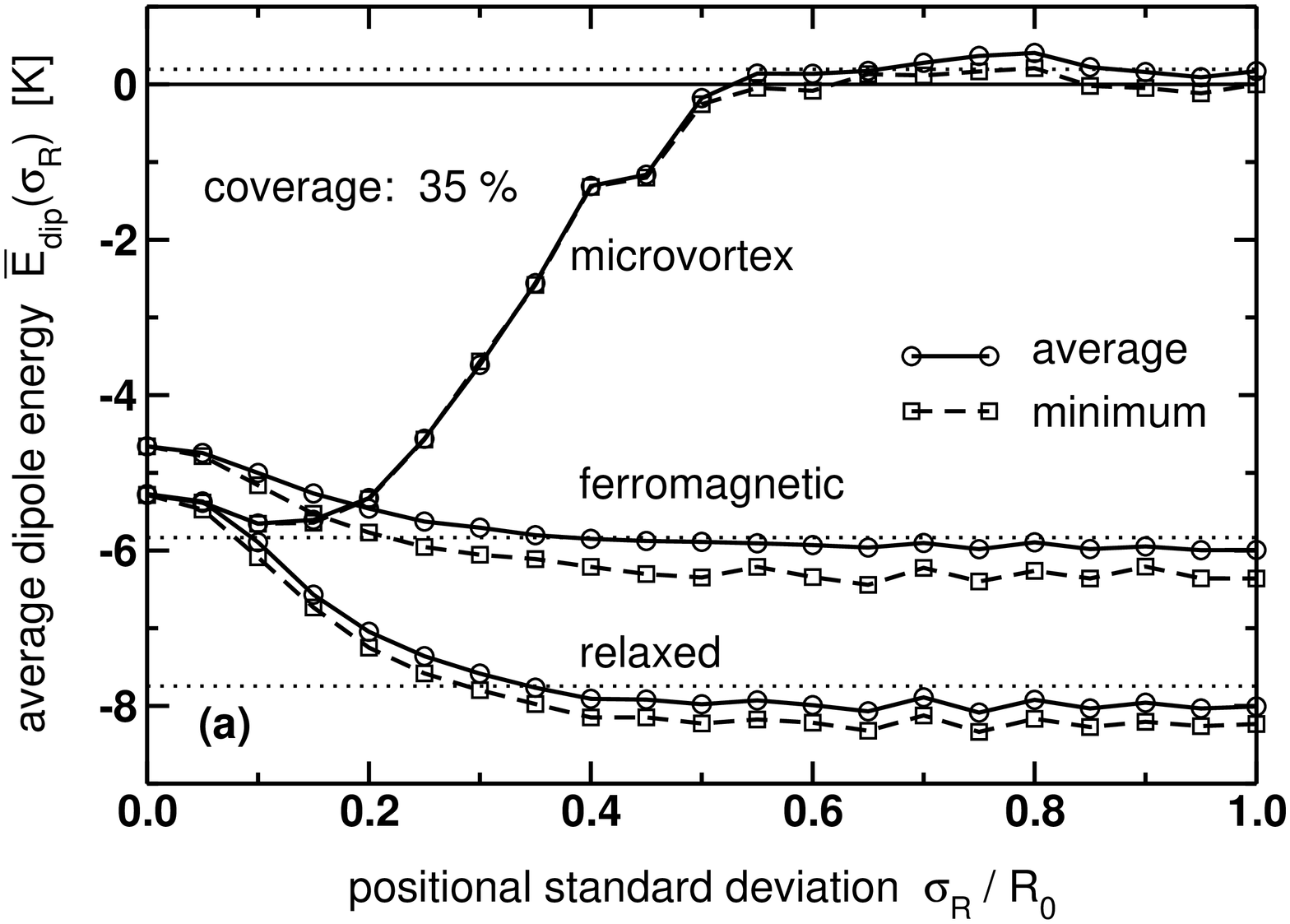}
\includegraphics[bb=60 1 600 740,angle=-90,clip,width=9cm]
        {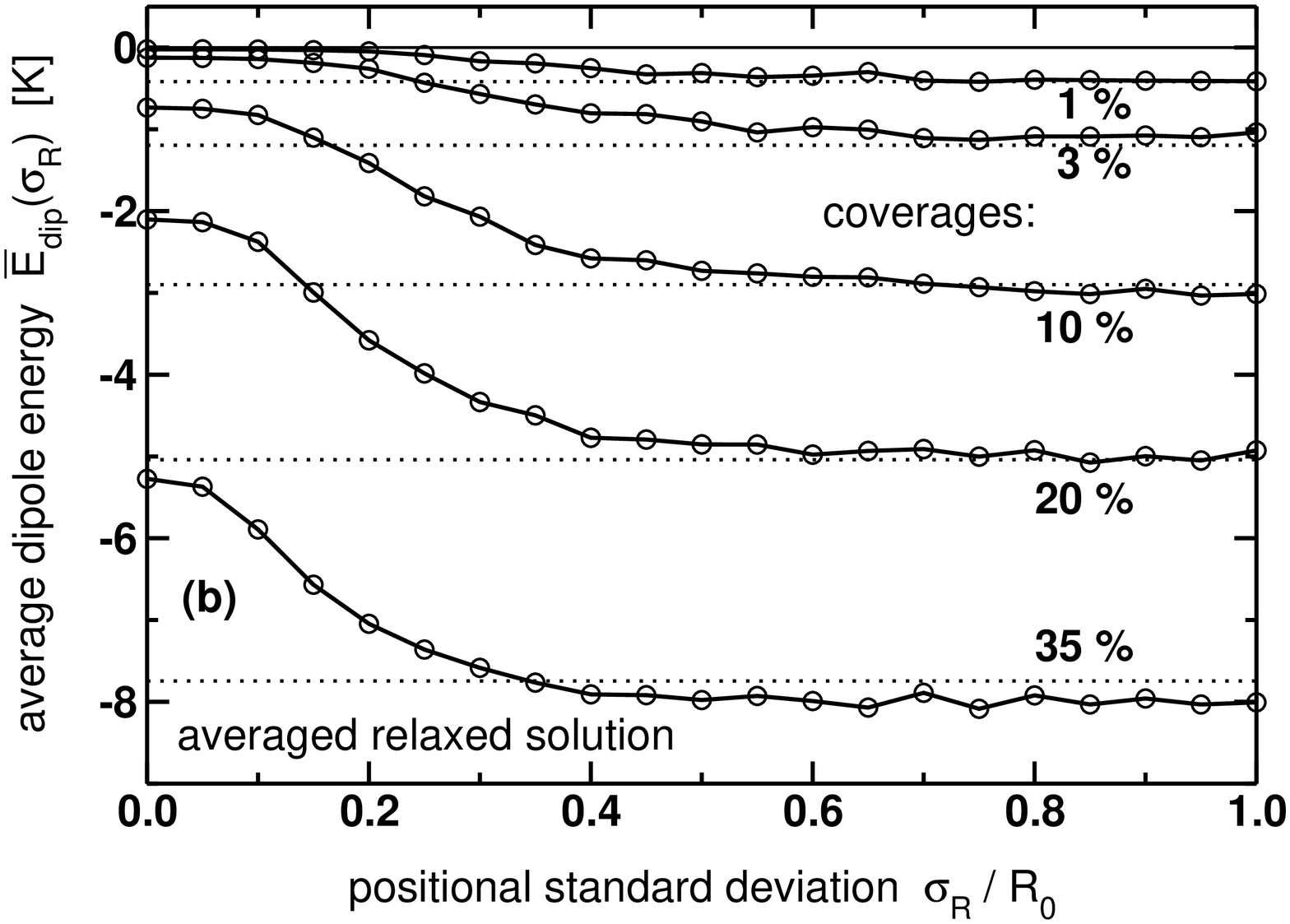}\\
\hspace*{4.5cm}\includegraphics[bb=60 1 600 740,angle=-90,clip,width=9cm]
        {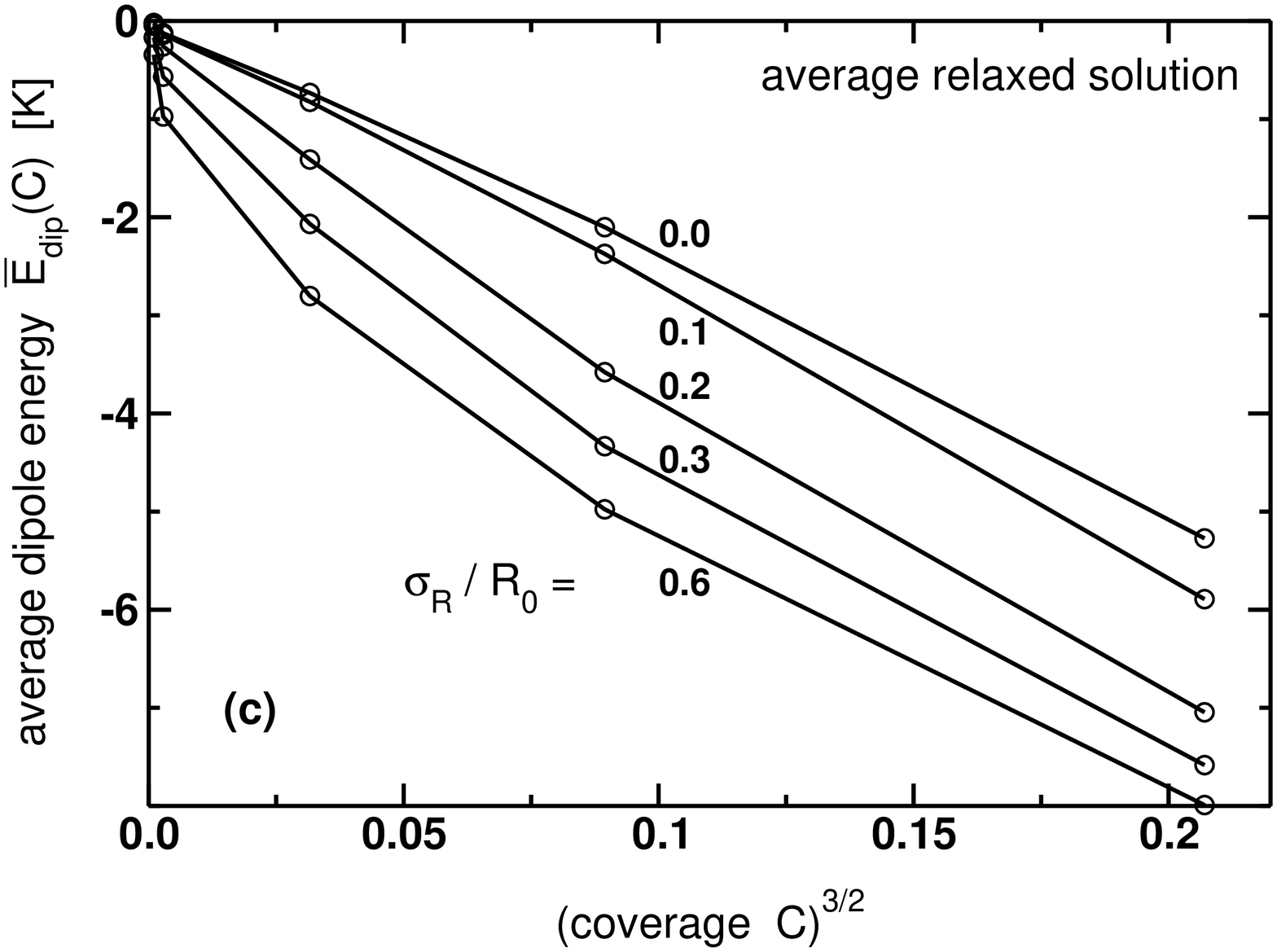}
\caption{
(a) Average dipole energy $\overline E_\mathrm{dip}(\sigma_R)$ 
per particle of a planar array of magnetic particles 
as function of the positional standard deviation $\sigma_R$. 
$R_0$ refers to the average interparticle distance, and 
the coverage amounts to $C=35$~\%. 
Displayed are $\overline E_\mathrm{dip}(\sigma_R)$ 
for the relaxed magnetic states, averaged over 40 different
initial conditions (full lines, circles), and the minimum energies 
(dashed lines, squares) which indicate the dispersion of the data. 
Results are also given for the ferromagnetic and microvortex magnetic 
arrangements, averaged over the in-plane angles. 
The dotted lines denote $\overline E_\mathrm{dip}$ for a random 
particle setup with same particle sizes and coverages. 
All results correspond to the average over 20 different realizations of 
the unit cell. \\ 
(b) $\overline E_\mathrm{dip}(\sigma_R,C)$ of relaxed magnetic 
arrangements as function of $\sigma_R$ for different coverages $C$ as 
indicated. The dotted lines denote 
$\overline E_\mathrm{dip}(C)$ for random particle setups. \\
(c) $\overline E_\mathrm{dip}(\sigma_R,C)$ of relaxed magnetic 
arrangements as function of the particle coverage $C$ for different 
$\sigma_R$ as indicated. We show 
$\overline E_\mathrm{dip}(C)$ as function of $C^{-3/2}$ which should
yield a linear behaviour from a simple scaling estimate. 
}\end{figure} 

Already for moderate disorder $\sigma_R/R_0=0.10$ the character of the 
energy distribution changes strongly. An almost  
\textit{symmetric,} Gaussian-like energy distribution is obtained around the 
average dipole energy $\overline E_\mathrm{dip}$, see figure~2(b).  
The number of metastable states has increased remarkedly. In fact, 
out of the considered 20000 random trials no single  
state is reached twice after relaxation. This is 
in particular true for the low-energy states which 
are obtained with a very small 
frequency [see the encircled region in figure~2(b)]. 
For an even stronger disorder the small peak in the frequency
distribution for the low-energy states disappears completely. 
Since the total number of metastable states increases strongly 
with increasing disorder, the corresponding catchment areas 
decreases dramatically. 
The obtained large number of metastable states is consistent with
experiments on ferrofluids, which show that 
after application and removal of an external magnetic field 
the same magnetic arrangement is seldomly reached for a second time 
\cite{Schwa}. 

Furthermore, from figure~2 one observes clearly that 
the average magnetic dipole energy $\overline E_\mathrm{dip}$ 
decreases with increasing positional standard deviation $\sigma_R$. 
In figure~3 we present results for $\overline E_\mathrm{dip}(\sigma_R,C)$ 
per particle as function of $\sigma_R$ and for different particle 
coverages $C$. A square lattice corresponds to $\sigma_R=0$. 
Results for random particle setups are also shown. 
$\overline E_\mathrm{dip}(\sigma_R)$ is calculated for the 
relaxed solutions as well as for the ferromagnetic and 
microvortex magnetic states averaged over the corresponding in-plane 
angles $\phi_\mathrm{fm}$ and $\phi_\mathrm{mv}$. The obtained 
minimum values of $\overline E_\mathrm{dip}(\sigma_R)$ are also 
displayed in order to illustrate the spread of the 
results around the average. 
Whereas the energy distributions shown in figure~2 are  
determined from a \textit{single} realization of the unit cell, here  
$\overline E_\mathrm{dip}(\sigma_R)$ is in addition 
averaged over 20 different realizations. 

In figure~3(a), $\overline E_\mathrm{dip}(\sigma_R)$ is shown for 
a coverage $C=35$~\%. With increasing $\sigma_R$ the average dipole energy 
\textit{decreases,} i.e., the respective magnetic binding energy 
\textit{increases} with \textit{increasing disorder}
\cite{Jen01}. For $\sigma_R/R_0\gtrsim0.5$, 
$\overline E_\mathrm{dip}(\sigma_R)$ approaches a constant 
value which corresponds, within the numerical dispersion of the data, to 
the average dipole energy for a \textit{random} particle array. 
The decrease of the average energy is caused by the nonlinear 
dependence of the
dipole interaction with respect to the interparticle distance
\cite{Jen01}. In fact, once disorder is introduced, 
the increase of $\overline E_\mathrm{dip}$ for 
enlarged distances $r_{ij}$ between some particle 
pairs  is more  than counterbalanced by a corresponding 
decrease for smaller  distances between  other pairs of particles. 
A similar behaviour is obtained for the ferromagnetic
arrangement, albeit with a larger $\overline E_\mathrm{dip}(\sigma_R)$. 
In contrast, for the MV state the average energy 
exhibits a minimum as function of $\sigma_R$ at $\sigma_R/R_0\sim0.15$, 
and approaches $\overline E_\mathrm{dip}=0$ with increasing $\sigma_R$. 
Relaxation in a disordered particle arrangement, see figure~1, is thus
crucial to the disorder induced reduction of 
$\overline E_\mathrm{dip}(\sigma_R)$. Let us recall that 
the FM and MV arrangements usually do not 
correspond to local energy minima for $\sigma_R>0$. 

In figure~3(b) we show $\overline E_\mathrm{dip}(\sigma_R)$ of the 
relaxed solutions and for  different particle coverages $C$. 
First of all one observes that the overall dependence of 
$\overline E_\mathrm{dip}$ on $\sigma_R$ is not significantly 
affected by $C$. Increasing the interparticle spacing $R_0$ decreases 
the magnitude of the average dipole energy, which should scale in
principle as $\overline E_\mathrm{dip}\propto R_0^{-3}\propto C^{3/2}$. 
The dependence of $\overline E_\mathrm{dip}$ on $C$ 
is depicted in figure~3(c) for the relaxed solutions. Indeed, 
the expected behaviour $\overline E_\mathrm{dip}\propto C^{3/2}$ 
is obtained for weak positional disorder 
$\sigma_R$. However, for strong disorder and for small coverages a 
different concentration dependence is observed. 
It seems that the strong magnetic noncollinearities of the magnetic 
configurations render the simple 
scaling expectation no longer applicable. It would be therefore 
interesting to investigate in detail the 
energy scaling as function of coverage $C$ especially 
in the limit of strongly disturbed arrays of magnetic particles. 

Note that the magnitude of the dipole 
energy is comparably small. This is a consequence of the disk-shaped 
particles assumed in our calculations. For compact sphere-shaped 
particles with the same radius $r_0$ as the disk-shaped ones, 
thus yielding the same surface coverage $C$, the corresponding 
dipole energy will be significantly enhanced due to the larger number 
of atoms $N=(r_0/a_0)^3$ per spherical particle. Moreover, 
similar as the positional disorder a particle-size dispersion yields 
noncollinear magnetic arrangements even if the particle centers form 
a periodic lattice. However, one observes that 
$\overline E_\mathrm{dip}$ does not vary strongly with increasing 
size dispersion \cite{PJJ}. 

\begin{figure}[t] 
\hspace*{1cm} \includegraphics[bb=60 1 600 740,angle=-90,clip,width=15cm]
        {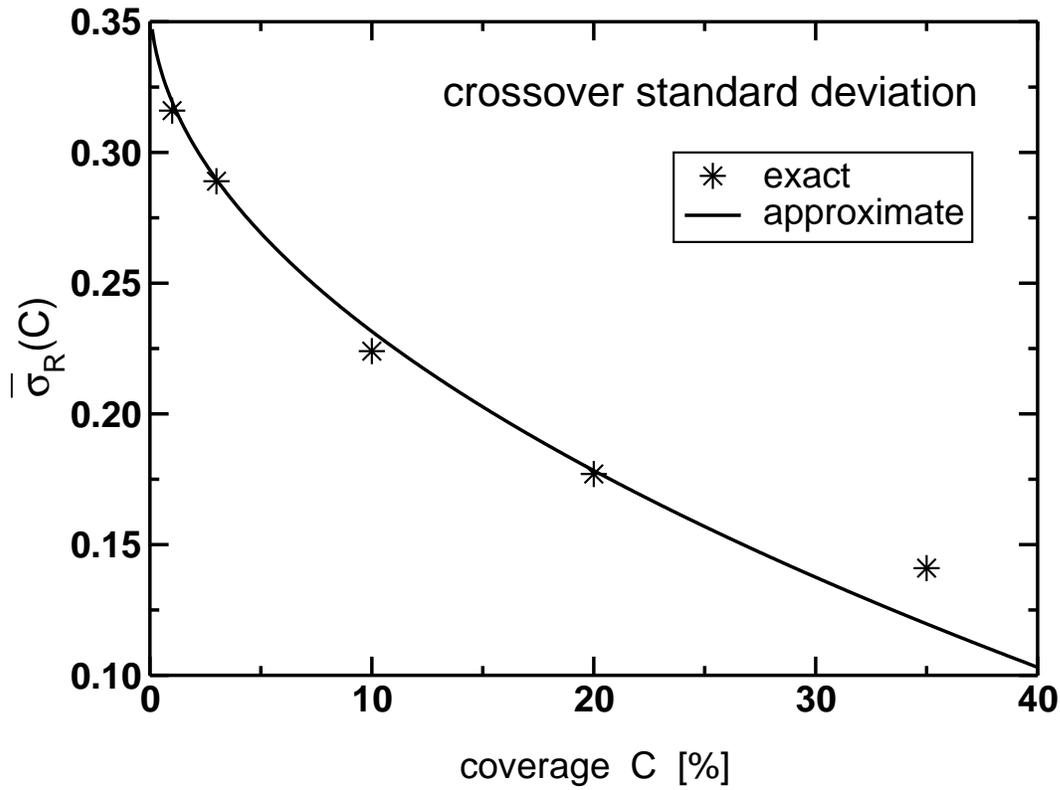}
\caption{
Crossover standard deviation $\overline\sigma_R$ as function of the
coverage $C$. For $\sigma_R>\overline\sigma_R$ 
the average dipole energy
per particle $\overline E_\mathrm{dip}(\sigma_R)$ of a disturbed square 
particle array converges to the one of a 
random particle setup. The symbols refer to the inflection points 
of $\overline E_\mathrm{dip}(\sigma_R)$ extracted from figure~3(b), 
and the full line 
to equation~\protect\ref{e3} using 0.72 as proportionality factor. 
}\end{figure} 
Furthermore, we discuss the crossover from a quasi-periodic to a 
random particle ensemble. In figure~4 we show the coverage dependence of 
the positional standard deviation $\overline\sigma_R$, above which the 
average energy $\overline E_\mathrm{dip}(\sigma_R)$ of a 
quasi-periodic particle arrangement reaches the limiting value 
of a random setup. One observes that $\overline\sigma_R$ decreases 
with increasing coverage $C$, as can qualitatively explained
by the following simple scaling consideration. 
The particles are scattered within a certain 
distance around the sites of the square lattice, the 
average scattering radius $R_{\sigma_R}$ can be estimated by 
$R_{\sigma_R}\simeq\sigma_R\,R_0$. For a given coverage $C$ 
the average dipole energy for a random particle setup is approximately
reached for the \textit{crossover standard deviation } 
$\overline\sigma_R$ for which $R_{\sigma_R}$ plus the particle 
radius equals half of the average interparticle distance, 
\begin{equation} 
R_{\overline\sigma_R}+r_0 \simeq R_0/2 \,. \label{e2} 
\end{equation}
Using $C=\pi(r_0/R_0)^2$, one obtains 
\begin{equation} 
\overline\sigma_R(C) \simeq \frac{1}{2}-\sqrt{\frac{C}{\pi}} \,. \label{e3} 
\end{equation} 
In figure~4, $\overline\sigma_R(C)$ is compared with the inflection 
points of $\overline E_\mathrm{dip}(\sigma_R)$ extracted from 
figure~3(b) for various coverages $C$. One observes that except for 
the largest considered coverage $C=35$~\% a satisfactory   
agreement is obtained, which supports the validity of the
previous scaling considerations. 
\begin{figure}[p] 
\hspace*{3cm}\includegraphics[bb=60 1 600 740,angle=-90,clip,width=12cm]
        {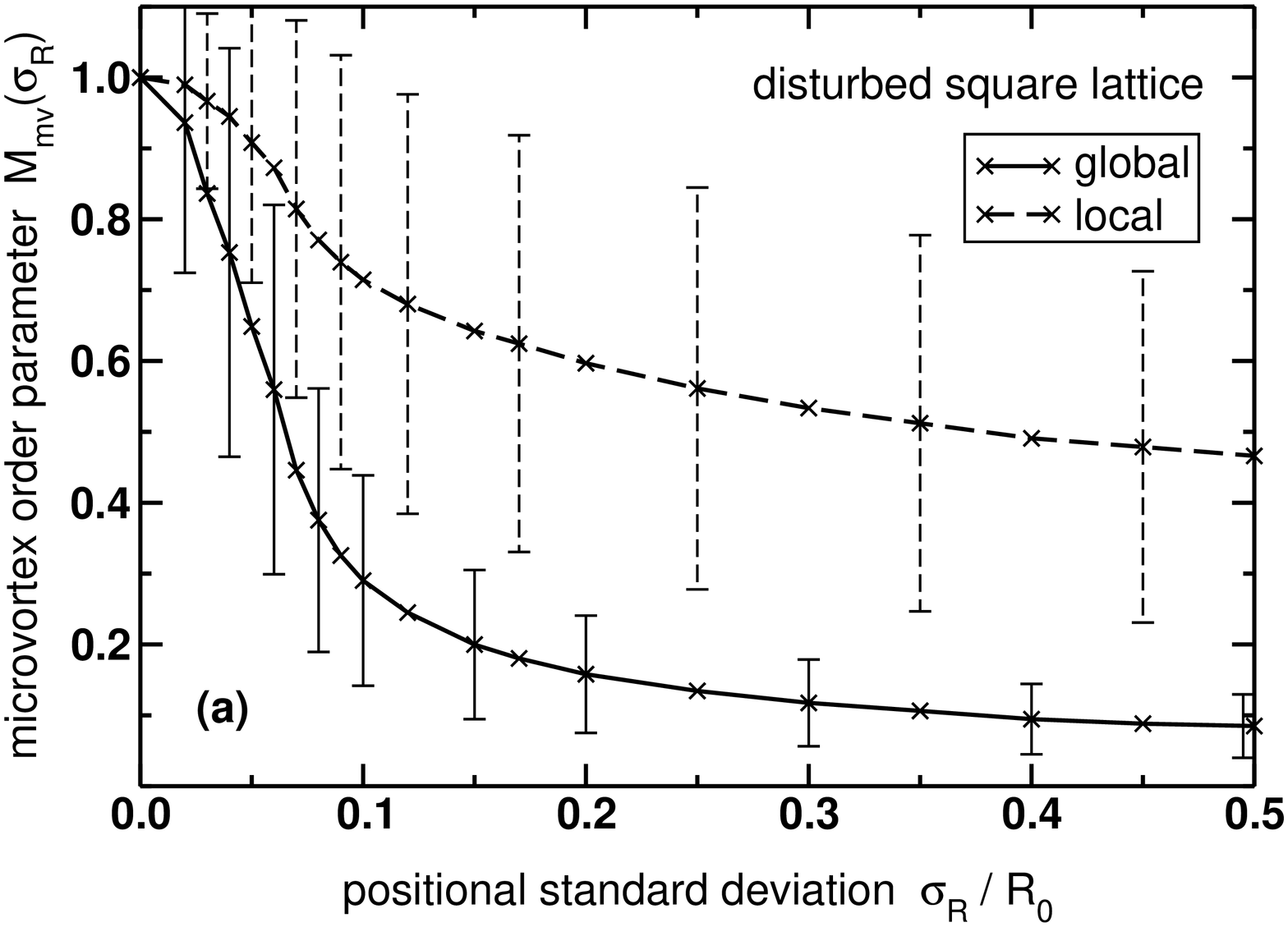}\\
\hspace*{3cm}\includegraphics[bb=60 1 600 740,angle=-90,clip,width=12cm]
        {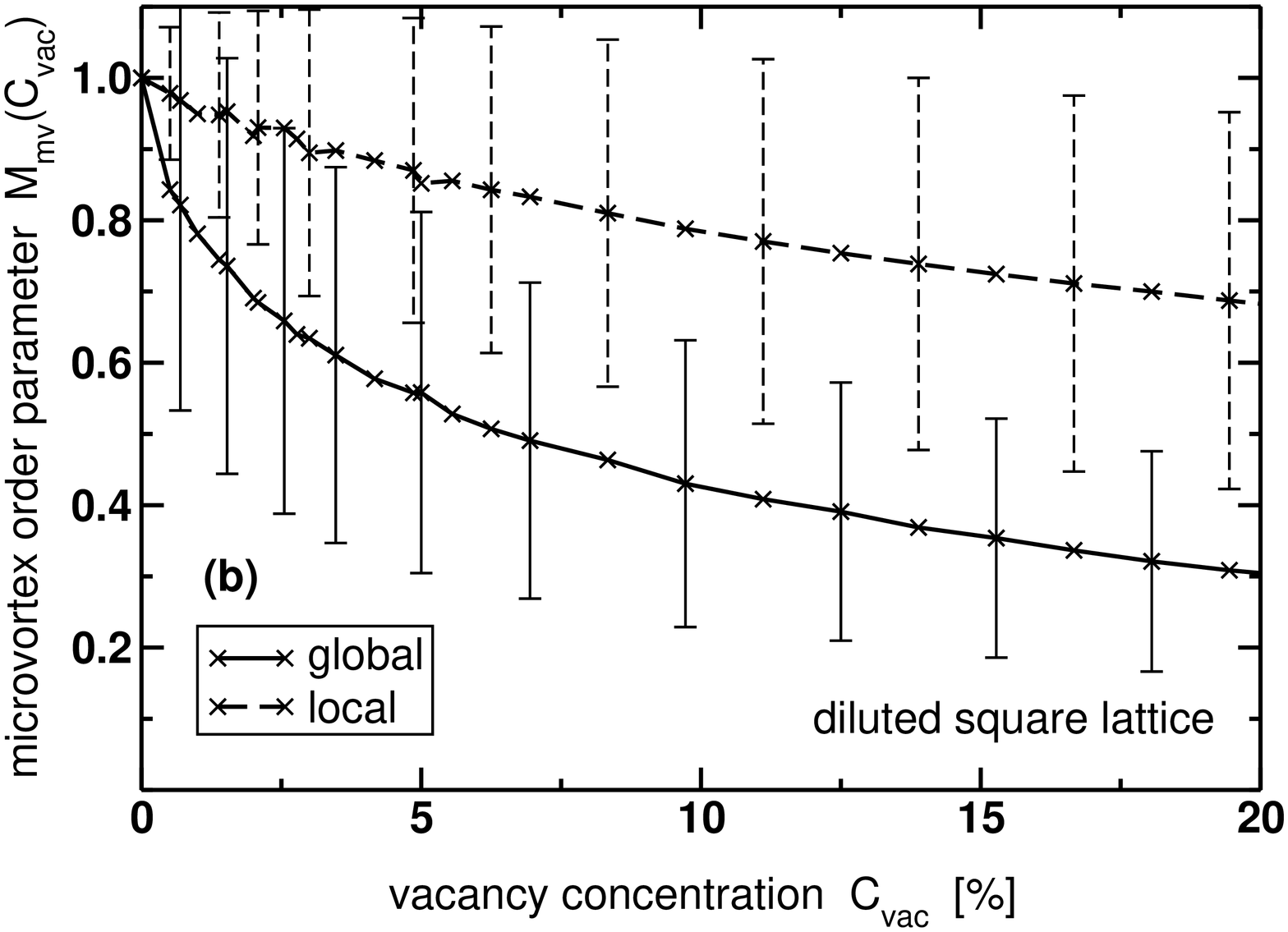}
\caption{Global and local microvortex order 
parameters $M_\mathrm{mv}^\mathrm{global}(\sigma_R)$ and 
$M_\mathrm{mv}^\mathrm{local}(\sigma_R)$ as function of the 
distortion of the particle array 
assuming (a) a disturbed square lattice with positional standard
deviation $\sigma_R$ and particle coverage $C=35$~\%, 
and (b) a square array with randomly distributed 
vacancies with concentration $C_\mathrm{vac}$.
These order parameters are averaged over many different initial
arrangements and different realizations of the unit cell, in addition 
$M_\mathrm{mv}^\mathrm{local}(\sigma_R)$ is averaged over all 
four-particle plaquettes in the unit cell. The resulting 
standard deviation is indicated by the error bars. 
}\end{figure} 

Results for the global and local microvortex order 
parameters $M_\mathrm{mv}^\mathrm{global}(\sigma_R)$ and 
$M_\mathrm{mv}^\mathrm{local}(\sigma_R)$ are presented in figure~5, 
assuming two different types of disorder. In figure~5(a) we consider 
a disturbed square particle lattice as function of the
positional standard deviation $\sigma_R$ for a coverage $C=35$~\%, 
and in figure~5(b) we vary the concentration of vacancies in an 
otherwise periodic array. The rather large dispersions of the MV order 
parameters result from many different initial
arrangements and different realizations of the unit cell. 
Locally the MV order is preserved even for strong disorder. For example, 
for $\sigma_R/R_0=0.5$ which refers to an almost random particle
array, see figure~3(a), 
we obtain $M_\mathrm{mv}^\mathrm{local}\sim0.5$,  
and $M_\mathrm{mv}^\mathrm{local}\sim0.7$ for $C_\mathrm{vac}=20$~\%. 
In contrast, positional disorder has a much stronger effect on 
$M_\mathrm{mv}^\mathrm{global}$. Especially 
positional disturbances quickly destroy the long-range MV 
ordering in the square lattice. For instance, for a particle
coverage $C=35$~\% this occurs already for 
$\sigma_R/R_0\gtrsim 0.10$, which is consistent with the minimum of 
$\overline E_\mathrm{dip}(\sigma_R)$ of the MV state, see figure~3(a). 
\begin{figure}[p]
\hspace*{3cm}\includegraphics[bb=60 1 600 740,angle=-90,clip,width=10cm]
        {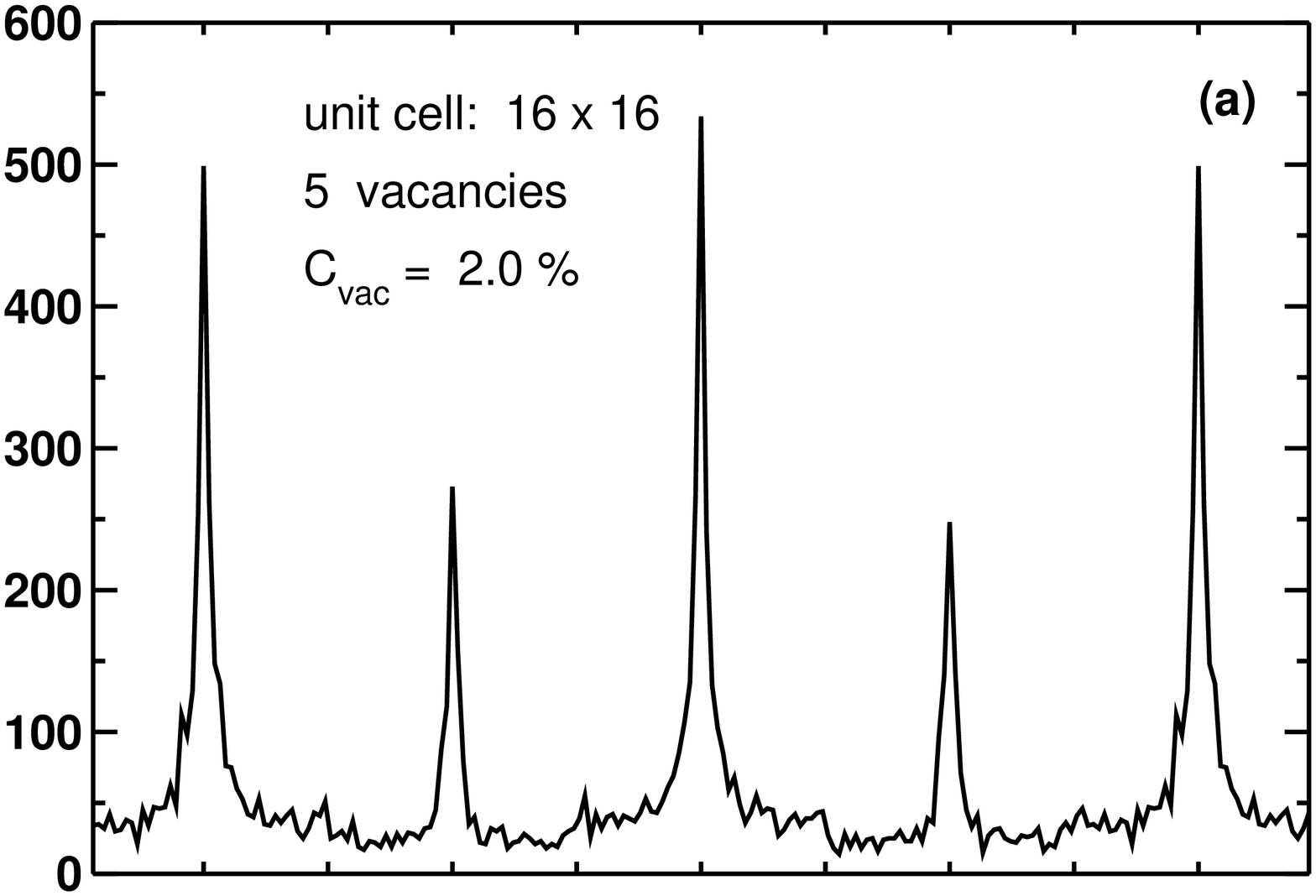}\\[-0.9cm]
\hspace*{3cm}\includegraphics[bb=60 1 600 740,angle=-90,clip,width=10cm]
        {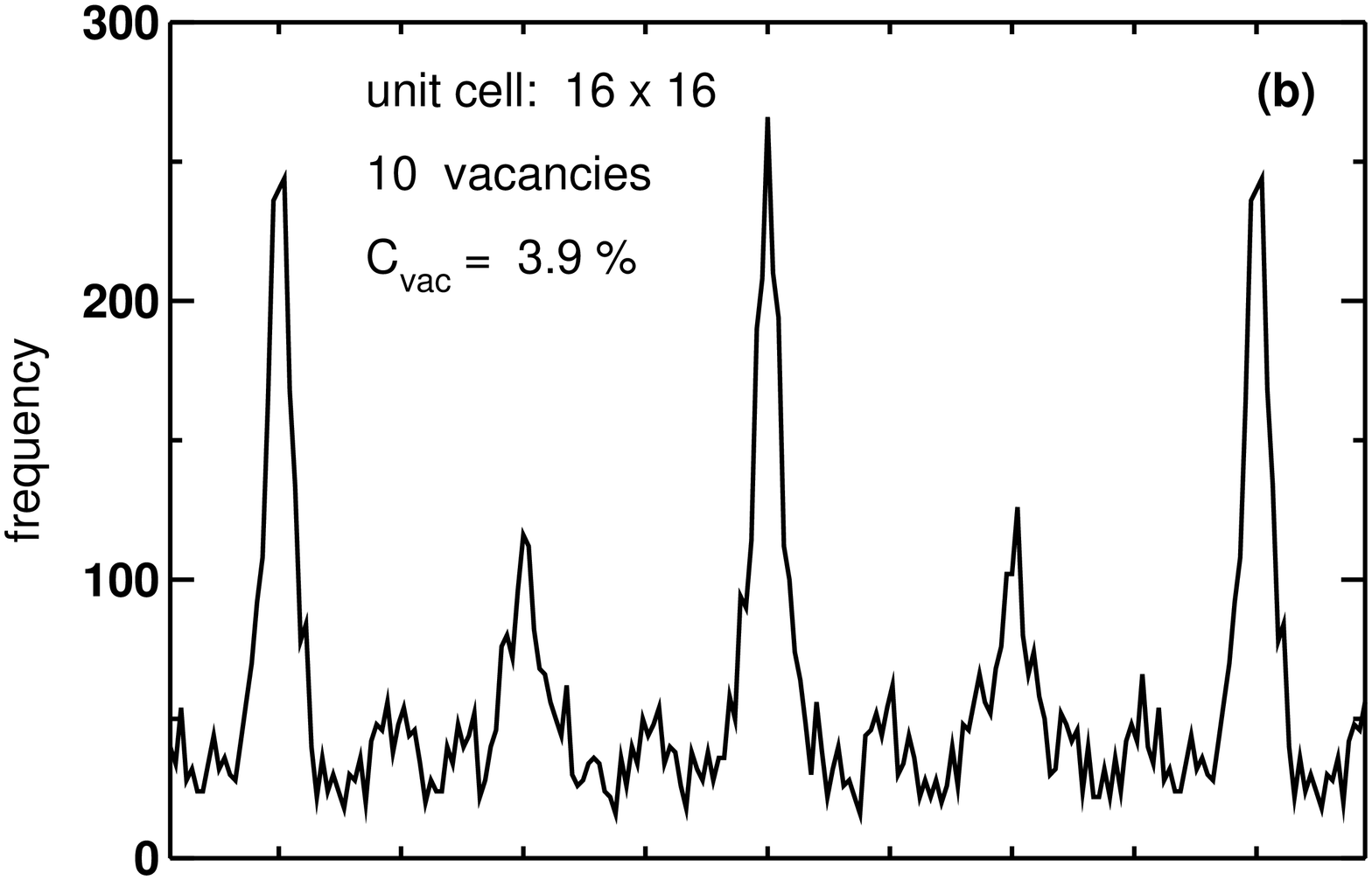}\\[-0.9cm]
\hspace*{3cm}\includegraphics[bb=60 1 600 740,angle=-90,clip,width=10cm]
        {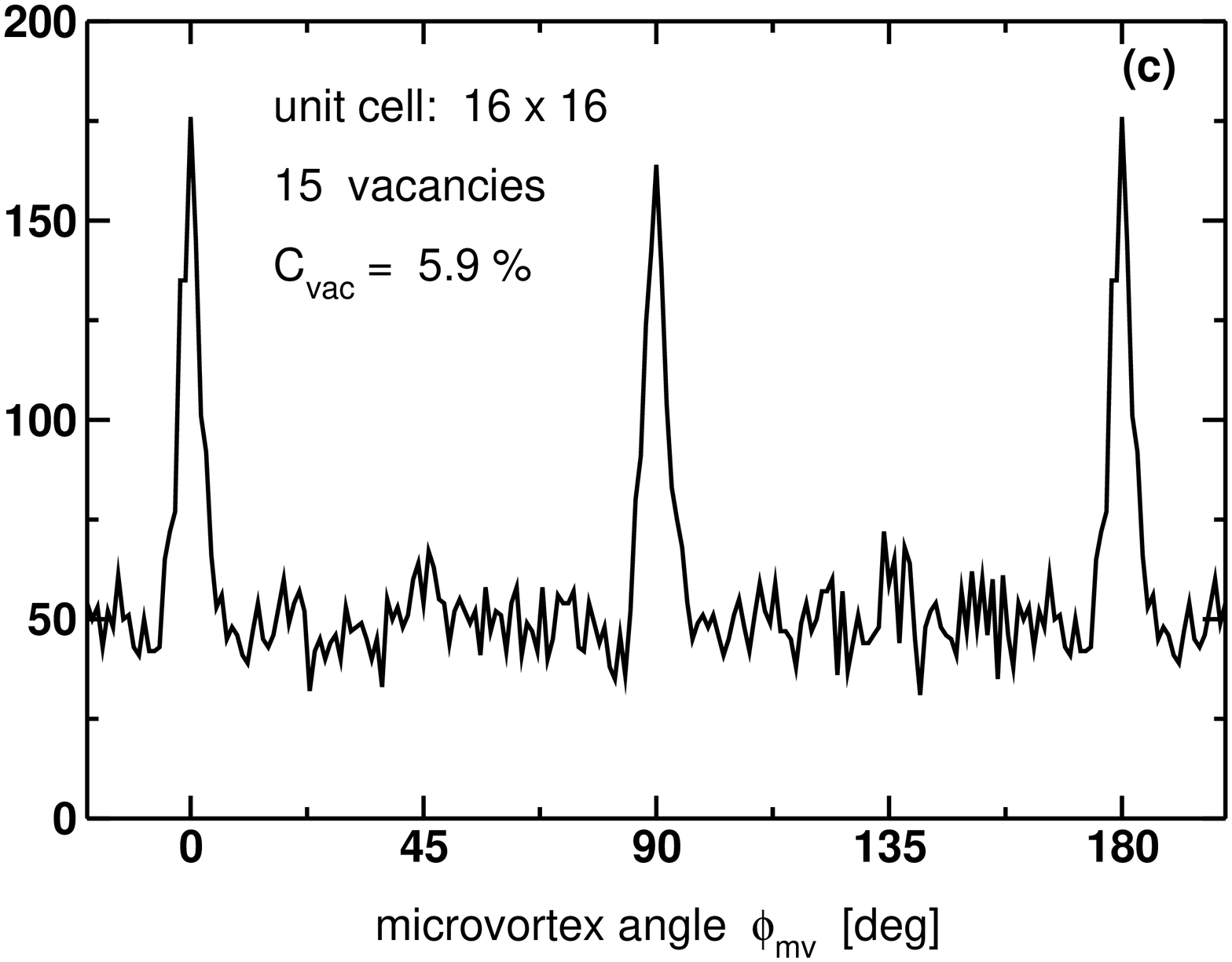}
\caption{
Angular distribution of the microvortex angle $\phi_\mathrm{mv}$ for 
different vacancy concentrations $C_\mathrm{vac}$ on a square lattice. 
(a) $C_\mathrm{vac}=2.0$~\% (5 vacancies per unit cell), 
(b) $C_\mathrm{vac}=3.9$~\% (10 vacancies), and 
(c) $C_\mathrm{vac}=5.9$~\% (15 vacancies). 
The size of the unit cell is $16\times16$ particles, 1000 different
realizations of the unit cell are performed. Due to the mirror
symmetry we show the angular distribution for 
$-20^\circ\le\phi_\mathrm{mv}\le200^\circ$. 
}\end{figure} 

Finally, we present results for the angular distribution of the 
microvortex angle $\phi_\mathrm{mv}$ for different degrees of
disorder. We consider first the effect of \textit{vacancies} in an otherwise 
periodic square particle lattice, as has been done previously 
by Prakash and Henley \cite{PrH90}. According to these
authors the preferred MV angles for a dipole-coupled square spin
lattice with a small amount of vacancies correspond to the diagonal 
directions. Indeed, as can be seen from figure~6, a strong frequency 
of the diagonal MV angles $\phi_\mathrm{mv}=45^\circ$ and $135^\circ$ 
is obtained for low vacancy concentrations $C_\mathrm{vac}$. 
The frequencies of these states decrease with 
increasing $C_\mathrm{vac}$, and vanish for 
$C_\mathrm{vac}\gtrsim6$~-~7~\%. In addition to the diagonal states 
we also obtain a strong frequency of the MV angle along the axial
directions (columnar states) with MV angles 
$\phi_\mathrm{mv}=0^\circ,\;90^\circ,$ and $180^\circ$. 
These states are also present for larger vacancy concentrations. 
We have investigated whether the angular distributions are affected 
by the finite size of the unit cell, in particular whether 
this could induce the pronounced frequencies of the columnar 
states. If this is true, the ratio of the frequencies of 
$\phi_\mathrm{mv}=0^\circ$ and $\phi_\mathrm{mv}=90^\circ$ is 
affected by a variation of the aspect ratio of the rectangular unit cell. 
We found that for 
not too small $C_\mathrm{vac}$ the dependence of the calculated 
frequencies for columnar and diagonal states as function of 
size and shape of the unit cell is quite weak. 
Hence, our results indicate that the presence of a few vacancies 
induces a more complicated angular distribution of the 
MV angle. In any case, we would like to point out that at present
we cannot rule out completely the possibility 
that the pronounced frequencies of the columnar states might be caused 
by the finite size of the unit cell. 

Concerning other types of disorder, we observe that 
for a \textit{disturbed} square lattice characterized by 
the positional standard deviation $\sigma_R$ there is no 
particular preference of the diagonal states. 
For this type of disorder only the columnar 
states exhibit a pronounced frequency as compared to the rest of the 
MV angles. The frequency peaks for columnar states decrease for 
increasing disorder and vanish above $\sigma_R/R_0\sim0.20$ 
for the particle coverage $C=35$~\%.  
A marked dependence of the angular distribution on the size
and shape of the unit cell is not observed also for this type of disorder. 
To conclude this discussion, we would like to 
emphasize that the angular distributions presented in figure~6 
were all obtained for a \textit{nonrelaxed} MV state. 
We have also performed corresponding calculations for the 
\textit{relaxed} solutions, and have determined the MV states 
which are closest to the noncollinear relaxed magnetic arrangements. 
In contrast to the results shown in figure~6 the corresponding 
angular histograms for the relaxed states exhibit a strong dependence
on the size and shape of the unit cell. Therefore, further 
investigations would be necessary in order to clarify this matter. 

\section{Conclusion} \label{sec:conc} 
In the present study we have investigated the low-energy
properties of disordered planar arrays of magnetic nanoparticles
interacting by the dipole coupling. Different kinds of disorder 
have been considered. 
Already small deviations from a square particle lattice lift 
the continuous degeneracy of the microvortex ground state. 
A strongly noncollinear magnetic order appears 
in the nanostructure which destroys the MV arrangement, as can
be seen from the corresponding order parameter. 
We have shown that 
with an increasing disorder the energy distribution of the metastable
states changes qualitatively. Our results indicate that it should
be very difficult to identify the ground state energy 
and its magnetic configuration for strongly
inhomogeneous particle arrangements. This behaviour is typical for  
spin-glass systems or random magnets. A detailed investigation of the 
spin-glass hehaviour of the nanostructure is certainly worthlike. 
This would require the consideration of additional properties 
such as the nonlinear susceptibility and the time dependent 
correlation functions \cite{Luo91,Bin86}. 

It has been shown that the average magnetic dipole energy 
of an ensemble of magnetic particles decreases with increasing 
positional disorder. Moreover, 
the average dipole energy $\overline E_\mathrm{dip}$ of a 
quasi-periodic particle arrangement resembles the one of a 
random particle setup above the coverage-dependent crossover 
standard deviation $\overline\sigma_R(C)$. A simple scaling 
behaviour of $\overline\sigma_R(C)$ has been derived, which reproduces 
quite accurately the coverage dependence of $\overline\sigma_R(C)$.
Structural disorder and strong magnetic noncollinearity effects result in
a deviation from the straightforward scaling 
$\overline E_\mathrm{dip}\propto C^{3/2}$ 
of the average dipole energy as function of the coverage. 

For a square particle lattice with a small vacancy concentration we
found a preferred orientation of the microvortex angle
$\phi_\mathrm{mv}$ along the diagonal directions. This is in
agreement with the results obtained previously by Prakash and Henley 
\cite{PrH90}. The pronounced frequency of this state decreases with an
increasing vacancy concentration. However, we also find indications 
that the columnar states are particularly frequent. 
The difference with previous results could be related to the fact 
that isolated vacancies were considered in \cite{PrH90}, whereas 
in the present study the vacancy concentration cannot 
be chosen arbitrarily small. In contrast to the case of 
vacancy-induced disorder, for a disturbed square particle array 
characterized by the positional standard deviation $\sigma_R$ a 
pronounced frequency is found only for
the columnar but not for the diagonal states. 

The determination of $\overline E_\mathrm{dip}$ allows to 
distinguish whether a magnetic particle ensemble can be considered 
as a weakly or a strongly interacting 
system. In the first case the interactions can be treated as 
perturbations to the single-particle couplings, whereas in the latter
case they have to be considered explicitly. The present study can be 
easily extended in order to take into account single-particle 
anisotropies with distributions of their magnitudes and easy axes.
The importance of collectively ordered magnetic states 
for strongly interacting particle systems has been pointed out. 
An increasing average binding energy $|\overline E_\mathrm{dip}|$ favours 
magnetic ordering and should cause an increase of its critical spin-glass 
temperature. In this context the effect of disorder in the particle 
ensemble on the critical temperature is of considerable interest. 
Also the magnetic relaxation will depend 
sensitively on the degree of disorder in such nanostructured systems. 
Finite temperature effects can be introduced in the 
framework of a mean field approximation or by performing Monte Carlo 
simulations. Magnetic hysteresis loops and susceptibilities can be derived 
by applying magnetic fields with different directions and strengths. 

\ack 

The authors acknowledge support from CNRS (France) and from the EU GROWTH 
project AMMARE (contract number G5RD-CT-2001-00478). We thank P.\ Politi
for fruitful discussions and for sending us his work prior to 
publication. 

\appendix
\section{Dipole interaction between magnetic particles with a 
finite size}
In this Appendix we determine the leading correction to the 
magnetic dipole interaction, see equation~\ref{e1}, beyond the 
point-dipole sum. The finite extension of the 
particles for the most general case of arbitrary sizes and shapes is  
taken into account. For a hexagonal lattice the finite particle size
has been considered already by Politi and Pini \cite{Pol02}. First, 
the interaction between a particle pair is calculated. Then a particle 
ensemble with an infinite lateral extension is modeled
by means of a rectangular unit cell with periodic boundary conditions. 

\subsection*{The dipole-quadrupole correction} 
Consider a particle $i$ containing $N_i$ atoms on lattice sites $k$ 
with position vectors $\bi{r}_{ik}=\bi{r}_{i0}+\bi{r}_k$. 
The center of gravity of this particle is given by 
\begin{equation} 
\bi{r}_{i0}= (x_{i0},y_{i0},z_{i0}) = 
\frac{1}{N_i}\sum_{k\in i}\,\bi{r}_{ik} = 
\frac{1}{N_i}\sum_{k\in i}\,(x_{ik},y_{ik},z_{ik}) \,. \label{e4}
\end{equation}
The finite size of particle $i$ is taken into account by 
the quadratic deviations with respect to its center, 
\begin{equation}
\la x_i^2\ra = \frac{1}{N_i}\sum_{k\in i}\,x_k^2  =
\frac{1}{N_i}\sum_{k\in i}\,(x_{ik}-x_{i0})^2\,, \label{e5}
\end{equation} 
and similarly for $\la y_i^2\ra$ and $\la z_i^2\ra$.  
These quantities depend on the size and shape  of
the particle and have the dimension of an area.  
Furthermore, we define $\varepsilon_{ik}$ by 
\begin{equation}
r_{ik}^2 = (x_{i0}+x_k)^2+(y_{i0}+y_k)^2+(z_{i0}+z_k)^2 
=r_{i0}^2\,(1+\varepsilon_{ik}) \,, \label{e6} 
\end{equation}
hence, 
\begin{equation} \varepsilon_{ik} = \frac{1}{r_{i0}^2}\Big(
2\,x_{i0}\,x_k+2\,y_{i0}\,y_k+2\,z_{i0}\,z_k+x_k^2+y_k^2+z_k^2
\Big)\;, \label{e7} \end{equation}
with $r_{i0}^2=x_{i0}^2+y_{i0}^2+z_{i0}^2$. 
The factor $(1+\varepsilon_{ik})$ appearing in the denominator 
of the dipole interaction energy, see equation~\ref{e1}, is expanded to 
second order in $x_k$, $y_k$, and $z_k$ as 
\begin{eqnarray}
\fl \frac{1}{(1+\varepsilon_{ik})^{5/2}} \,\simeq\, 1-\frac{5}{2}\,
\varepsilon_{ik}+\frac{35}{8}\,\varepsilon^2 \,\simeq\, \nonumber \\ 
\fl 1-\frac{5}{2\,r_{i0}^2}\bigg(
2\,x_{i0}\,x_k+2\,y_{i0}\,y_k+2\,z_{i0}\,z_k+x_k^2+y_k^2+z_k^2 \bigg)
+\frac{35}{2\,r_{i0}^4}\,\bigg(x_{i0}\,x_k+y_{i0}\,y_k+z_{i0}\,z_k
\bigg)^2\;. \label{e8} \end{eqnarray} 
Now all different sums over the atomic sites $k$ of particle $i$ 
are performed up to this order. The non-vanishing terms are listed 
in the following equations~\ref{e9}--\ref{e18}. 
\begin{eqnarray}
\fl \sum_{k\in i}\frac{1}{r_{ik}^5} = \sum_{k\in i}
\frac{1}{r_{i0}^5(1+\varepsilon)^{5/2}} \nonumber \\  
\fl \hspace*{1.5cm} \simeq
\frac{N_i}{r_{i0}^5}+\frac{5}{2}\frac{N_i}{r_{i0}^7}\bigg[
\Big(7\,\frac{x_{i0}^2}{r_{i0}^2}-1\Big)\la x_i^2\ra
+\Big(7\,\frac{y_{i0}^2}{r_{i0}^2}-1\Big)\la y_i^2\ra +
\Big(7\,\frac{z_{i0}^2}{r_{i0}^2}-1\Big)\la z_i^2\ra\bigg]\,, 
\label{e9} \\
\fl \sum_{k\in i}\frac{x_{ik}}{r_{ik}^5} = \sum_{k\in i}
\frac{x_{i0}+x_k}{r_{i0}^5(1+\varepsilon)^{5/2}} \simeq 
\frac{N_i}{r_{i0}^5}\,x_{i0} +\frac{5}{2}\frac{N_i}{r_{i0}^7}\bigg[
\Big(7\,\frac{x_{i0}^3}{r_{i0}^2}-3\,x_{i0}\Big)\la x_i^2\ra \nonumber
\\ \fl \hspace*{1.5cm} 
+\Big(7\,\frac{x_{i0}\,y_{i0}^2}{r_{i0}^2}-x_{i0}\Big)\la
y_i^2\ra + \Big(7\,\frac{x_{i0}\,z_{i0}^2}{r_{i0}^2}-x_{i0}\Big) \la
z_i^2\ra \bigg]\,, \label{e10} \\
\fl \sum_{k\in i}\frac{y_{ik}}{r_{ik}^5} = \sum_{k\in i}
\frac{y_{i0}+y_k}{r_{i0}^5(1+\varepsilon)^{5/2}} \simeq 
\frac{N_i}{r_{i0}^5}\,y_{i0} +\frac{5}{2}\frac{N_i}{r_{i0}^7}\bigg[
\Big(7\,\frac{x_{i0}^2\,y_{i0}}{r_{i0}^2}-y_{i0}\Big)\la x_i^2\ra
\nonumber \\ \fl \hspace*{1.5cm} 
+\Big(7\,\frac{y_{i0}^3}{r_{i0}^2}-3\,y_{i0}\Big)\la
y_i^2\ra + \Big(7\,\frac{y_{i0}\,z_{i0}^2}{r_{i0}^2}-y_{i0}\Big) \la
z_i^2\ra \bigg]\,, \label{e11} \\
\fl \sum_{k\in i}\frac{z_{ik}}{r_{ik}^5} = \sum_{k\in i}
\frac{z_{i0}+z_k}{r_{i0}^5(1+\varepsilon)^{5/2}} \simeq 
\frac{N_i}{r_{i0}^5}\,z_{i0} +\frac{5}{2}\frac{N_i}{r_{i0}^7}\bigg[
\Big(7\,\frac{x_{i0}^2\,z_{i0}}{r_{i0}^2}-z_{i0}\Big)\la x_i^2\ra
\nonumber \\ \fl \hspace*{1.5cm} 
+\Big(7\,\frac{y_{i0}^2\,z_{i0}}{r_{i0}^2}-z_{i0}\Big)\la y_i^2\ra +
\Big(7\,\frac{z_{i0}^3}{r_{i0}^2}-3\,z_{i0}\Big) \la z_i^2\ra
\bigg]\,, \label{e12} \\
\fl \sum_{k\in i}\frac{x_{ik}^2}{r_{ik}^5} = \sum_{k\in i}
\frac{x_{i0}^2+2x_{i0}\,x_k+x_k^2}{r_{i0}^5(1+\varepsilon)^{5/2}}
\simeq \frac{N_i}{r_{i0}^5}\Big(x_{i0}^2+\la
x_i\ra^2\Big) +\frac{5}{2}\frac{N_i}{r_{i0}^7}\bigg[
\Big(7\,\frac{x_{i0}^4}{r_{i0}^2}-5\,x_{i0}^2\Big) \la x_i^2\ra
\nonumber \\ \fl \hspace*{1.5cm} 
+\Big(7\,\frac{x_{i0}^2\,y_{i0}^2}{r_{i0}^2}-x_{i0}^2\Big) \la
y_i^2\ra + \Big(7\,\frac{x_{i0}^2\,z_{i0}^2}{r_{i0}^2}-x_{i0}^2\Big)
\la z_i^2\ra \bigg]\,, \label{e13} \\
\fl \sum_{k\in i}\frac{y_{ik}^2}{r_{ik}^5} = \sum_{k\in i}
\frac{y_{i0}^2+2y_{i0}\,y_k+y_k^2}{r_{i0}^5(1+\varepsilon)^{5/2}}
\simeq \frac{N_i}{r_{i0}^5}\Big(y_{i0}^2+\la
y_i\ra^2\Big) +\frac{5}{2}\frac{N_i}{r_{i0}^7}\bigg[
\Big(7\,\frac{x_{i0}^2\,y_{i0}^2}{r_{i0}^2}-y_{i0}^2\Big) \la x_i^2\ra
\nonumber \\ \fl \hspace*{1.5cm} 
+\Big(7\,\frac{y_{i0}^4}{r_{i0}^2}-5\,y_{i0}^2\Big)\la y_i^2\ra +
\Big(7\,\frac{y_{i0}^2\,z_{i0}^2}{r_{i0}^2}-y_{i0}^2\Big) \la z_i^2\ra
\bigg]\,, \label{e14} \\
\fl \sum_{k\in i}\frac{z_{ik}^2}{r_{ik}^5} = \sum_{k\in i}
\frac{z_{i0}^2+2z_{i0}\,z_k+z_k^2}{r_{i0}^5(1+\varepsilon)^{5/2}}
\simeq \frac{N_i}{r_{i0}^5}\Big(z_{i0}^2+\la
z_i\ra^2\Big) +\frac{5}{2}\frac{N_i}{r_{i0}^7}\bigg[
\Big(7\,\frac{x_{i0}^2\,z_{i0}^2}{r_{i0}^2}-z_{i0}^2\Big) \la x_i^2\ra
\nonumber \\ \fl \hspace*{1.5cm} 
+\Big(7\,\frac{y_{i0}^2\,z_{i0}^2}{r_{i0}^2}-z_{i0}^2\Big) \la
y_i^2\ra + \Big(7\,\frac{z_{i0}^4}{r_{i0}^2}-5\,z_{i0}^2\Big) \la
z_i^2\ra \bigg]\,, \label{e15} \\
\fl \sum_{k\in i}\frac{x_{ik}\,y_{ik}}{r_{ik}^5} = \sum_{k\in i}
\frac{x_{i0}\,y_{i0}+x_{i0}\,y_k+x_k\,y_{i0}+x_k\,y_k}
{r_{i0}^5(1+\varepsilon)^{5/2}}  \nonumber \\  \fl \hspace*{1.5cm} 
\simeq \frac{N_i}{r_{i0}^5}\,x_{i0}\,y_{i0}
+\frac{5}{2}\frac{N_i}{r_{i0}^7}\bigg[
\Big(7\,\frac{x_{i0}^3\,y_{i0}}{r_{i0}^2}-3\,x_{i0}\,y_{i0}\Big) \la
x_i^2\ra \nonumber \\ \fl \hspace*{1.5cm} 
+\Big(7\,\frac{x_{i0}\,y_{i0}^3}{r_{i0}^2}-3\,x_{i0}\,y_{i0}\Big) \la
y_i^2\ra+\Big(7\,\frac{x_{i0}\,y_{i0}\,z_{i0}^2}{r_{i0}^2}-x_{i0}\,
y_{i0}\Big) \la z_i^2\ra \bigg]\,, \label{e16} \\
\fl \sum_{k\in i}\frac{x_{ik}\,z_{ik}}{r_{ik}^5} = \sum_{k\in i}
\frac{x_{i0}\,z_{i0}+x_{i0}\,z_k+x_k\,z_{i0}+x_k\,z_k}
{r_{i0}^5(1+\varepsilon)^{5/2}}  \nonumber \\ \fl \hspace*{1.5cm} 
\simeq \frac{N_i}{r_{i0}^5}\,x_{i0}\,z_{i0}
+\frac{5}{2}\frac{N_i}{r_{i0}^7}\bigg[
\Big(7\,\frac{x_{i0}^3\,z_{i0}}{r_{i0}^2}-3\,x_{i0}\,z_{i0}\Big) \la
x_i^2\ra \nonumber \\ \fl \hspace*{1.5cm} 
+\Big(7\,\frac{x_{i0}\,y_{i0}^2\,z_{i0}}{r_{i0}^2}-x_{i0}\,z_{i0}\Big)
\la y_i^2\ra+\Big(7\,\frac{x_{i0}\,z_{i0}^3}{r_{i0}^2}-3\,x_{i0}\,
z_{i0}\Big) \la z_i^2\ra \bigg]\,, \label{e17} \\
\fl \sum_{k\in i}\frac{y_{ik}\,z_{ik}}{r_{ik}^5} = \sum_{k\in i}
\frac{y_{i0}\,z_{i0}+y_{i0}\,z_k+y_k\,z_{i0}+y_k\,z_k}
{r_{i0}^5(1+\varepsilon)^{5/2}}  \nonumber \\ \fl \hspace*{1.5cm} 
\simeq \frac{N_i}{r_{i0}^5}\,y_{i0}\,z_{i0} +\frac{5N_i}{2r_{i0}^7}\bigg[
\Big(7\,\frac{x_{i0}^2\,y_{i0}\,z_{i0}}{r_{i0}^2}-y_{i0}\,z_{i0}\Big)
\la x_i^2\ra \nonumber \\ \fl \hspace*{1.5cm} 
+\Big(7\,\frac{y_{i0}^3\,z_{i0}}{r_{i0}^2}-3\,y_{i0}\,z_{i0}\Big) \la
y_i^2\ra+\Big(7\,\frac{y_{i0}\,z_{i0}^3}{r_{i0}^2}-3\,y_{i0}\,
z_{i0}\Big) \la z_i^2\ra \bigg]\,. \label{e18}
\end{eqnarray}

A simple extension of the previous considerations allows to calculate 
to the same order the corresponding sums involved in the interaction
between two extended particles $i$ and $j$ with sizes $N_i$ and 
$N_j$. Let $\bi{r}_{ij0}= (x_{ij0},y_{ij0},z_{ij0})$ 
denote the relative position vector between the particle centers.
The non-vanishing sums in the interaction energy
are given by equations~\ref{e9}--\ref{e18} after the replacements 
$N_j\,\to\,N_i\,N_j$, $r_{j0}\,\to\,r_{ij0}$, $x_{j0}\,\to\,x_{ij0}$, 
$\la x_j^2\ra\,\to\,\la x_i^2\ra+\la x_j^2\ra$, etc., have been performed. 
Concerning the summation over all atoms $k'$ of particle $j$,
one obtains, for example, for equation~\ref{e9}, 
\begin{eqnarray}
\fl \sum_{k\in i}\sum_{k'\in j}
\frac{1}{|\bi{r}_{ik}-\bi{r}_{jk'}|^5} \simeq
N_i\,N_j\bigg[\frac{1}{r_{ij0}^5} +\frac{5}{2\,r_{ij0}^7}
\Big(7\,\frac{x_{ij0}^2}{r_{ij0}^2} -1\Big)\Big(\la x_i^2\ra+\la
x_j^2\ra \Big) \nonumber \\ \fl \hspace*{1.5cm} +\frac{5}{2\,r_{ij0}^7}
\Big(7\,\frac{y_{ij0}^2}{r_{ij0}^2}-1\Big)\Big(\la y_i^2\ra+\la
y_j^2\ra  \Big) + \frac{5}{2\,r_{ij0}^7}
\Big(7\,\frac{z_{ij0}^2}{r_{ij0}^2}-1\Big)\Big(\la z_i^2\ra+\la
z_j^2\ra  \Big) \Bigg] \;. \label{e19}
\end{eqnarray}
Thus, within this expansion the dipole interaction between 
particles $i$ and $j$ is expressed in terms of 
the distance $r_{ij0}$ between their centers and the 
quadratic deviations $\la x_i^2\ra$, $\la x_j^2\ra$, etc., 
characterizing the sizes and shapes of the two particles. 
The point-dipole sum is recovered by setting $\la x_i^2\ra=0$,
etc. The correction to the point-dipole sum 
is a dipole-quadrupole interaction, being of the order 
$(\la x_i^2\ra+\la x_j^2\ra)/r_{ij0}^2$, i.e.,   
the ratio of the particle extensions and the square of the 
interparticle distance. Evidently, the effect of 
finite particle sizes becomes more 
important the closer the particles are located. This is the case, 
for example, for a densely packed 3D ferrofluid or for a layered 
nanostructured particle ensemble with a large surface coverage.  

In figure~7 we compare the exact dipole interaction 
energy $E_\mathrm{dip}$ with the results obtained by using 
the point-dipole approximation with and without the dipole-quadrupole 
correction. $E_\mathrm{dip}(d)$ for two square-shaped particles is shown 
as function of the distance $d$ between their centers. 
Notice that the correction for nearby particles amounts to 
$\sim25$~\% of the point-dipole sum in this case. 
For simple particle geometries (e.g., disks, cubes, or spheres)
the quadratic deviations $\la x_i^2\ra$, etc., can be calculated 
analytically. In case of spheres the dipole-quadrupole 
correction vanishes since a sphere has no quadrupole moment \cite{YGP}.  
\begin{figure}[t]
\hspace*{1cm} \includegraphics[bb=60 0 600 750,angle=-90,clip,width=15cm]
        {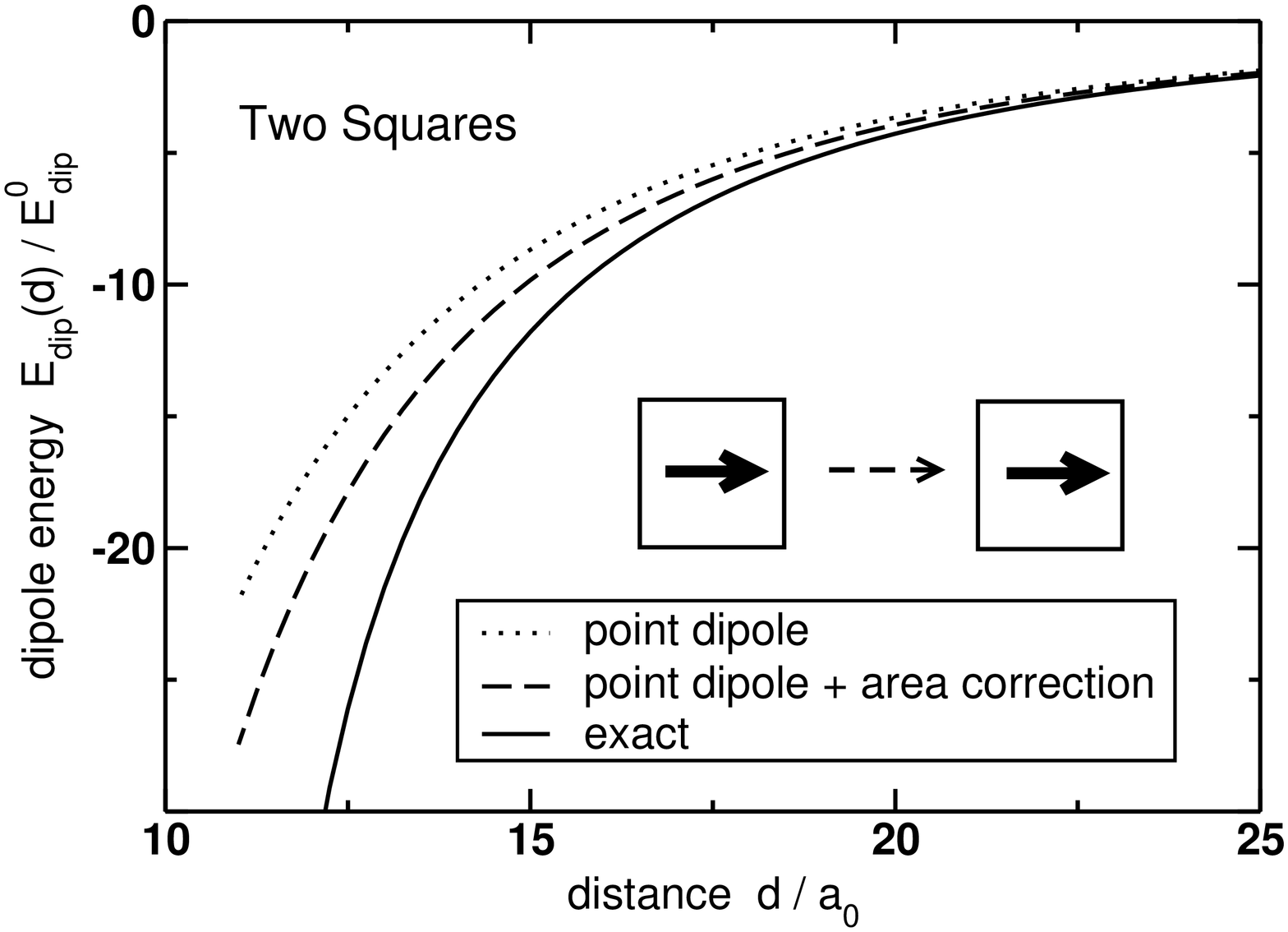}
\caption{
Comparison between the point-dipole approximation, the 
point-dipole approximation augmented by the dipole-quadrupole correction, 
and the exactly calculated dipole interaction $E^\mathrm{dip}(d)$.  
We consider two square-shaped flat (001) magnetic particles, consisting 
of $11\times11=121$ atomic magnetic moments each.  
$E_\mathrm{dip}(d)$, in units of
$E_\mathrm{dip}^0=\mu_\mathrm{at}^2/a_0^3$, is shown as function of the 
distance $d$ between the particle centers in units of the lattice
constant $a_0$, where $\mu_\mathrm{at}$ is the atomic magnetic moment. 
Note that for $d/a_0=10$ the particles touch each other. The thick 
arrows illustrate the 
magnetization directions, which are oriented head-to-tail. 
}\end{figure} 

\subsection*{The extended planar system} 
Let us now consider a planar particle ensemble infinitely extended in
the $xy$ plane. The unit cell has the size $L_x\times L_y$ and 
consists $n$ particles.  
Periodic boundary  conditions are applied laterally, whereas the 
vertical extension along the $z$-direction is finite. 
The dipole interaction sum runs over all particle pairs within 
the same and between different unit cells. 
The position vectors connecting two particle centers are given by 
$\bi{r}_{ij0}=(x_{ij0}+l_x\,L_x,y_{ij0}+l_y\,L_y,z_{ij0})$, with
$l_x$ and $l_y$ integers. For such a periodic planar 
system an Ewald summation technique can be applied 
by using a rapidly convergent 2D lattice summation \cite{Jen97}. 
The following general sums over the unit cells need to be considered,
\begin{equation}
T_{ij}^{\alpha\beta\gamma}=\sum_{l_x,l_y=-\infty}^\infty{}\hspace{-0.4cm}'
\hspace{0.4cm}\frac{x_{ij0}^\beta\,y_{ij0}^\gamma}{r_{ij0}^\alpha}\,. 
\label{e20} \end{equation} 
The prime at the sum indicates that the term with
$r_{ij0}=0$ is omitted. The different lattice sums appearing
in equations~\ref{e9}--\ref{e18} can be obtained by an appropriate
choice of the positive integers
$\alpha$, $\beta$, and $\gamma$.  In general one has
to distinguish between the  cases $z_{ij0}=0$ and $z_{ij0}\neq0$.

The dipole field $\bi{B}_i^\mathrm{dip}$ of particle $i$, see 
equation~\ref{e1}, has the components 
\begin{eqnarray} 
\fl B_i^{x,\mathrm{dip}}=\mu_{at}\,\sum_j N_j \hspace{-0.2cm} 
\sum_{l_x,l_y=-\infty}^\infty{}\hspace{-0.4cm}'\hspace{0.3cm}
\frac{1}{r^5}\bigg[(2x_{ij0}^2-y_{ij0}^2-z_{ij0}^2)\,m_j^x
+3\,x_{ij0}y_{ij0}\,m_j^y+3\,x_{ij0}z_{ij0}\,m_j^z\bigg]\,, \label{e21} \\ 
\fl B_i^{y,\mathrm{dip}}=\mu_{at}\,\sum_j N_j \hspace{-0.2cm}
\sum_{l_x,l_y=-\infty}^\infty{}\hspace{-0.4cm}'\hspace{0.3cm}
\frac{1}{r^5}\bigg[(2y_{ij0}^2-x_{ij0}^2-z_{ij0}^2)\,m_j^y
+3\,x_{ij0}y_{ij0}\,m_j^x+3\,y_{ij0}z_{ij0}\,m_j^z\bigg]\,, \label{e22} \\
\fl B_i^{z,\mathrm{dip}}=\mu_{at}\,\sum_j N_j \hspace{-0.2cm} 
\sum_{l_x,l_y=-\infty}^\infty{}\hspace{-0.4cm}'\hspace{0.3cm}
\frac{1}{r^5}\bigg[(2z_{ij0}^2-x_{ij0}^2-y_{ij0}^2)\,m_j^z
+3\,x_{ij0}z_{ij0}\,m_j^x+3\,y_{ij0}z_{ij0}\,m_j^y\bigg]\,. \label{e23}
\end{eqnarray} 
The unit vector $\bi{m}_j=\bi{M}_j/N_j=(m_j^x,m_j^y,m_j^z)=
(\sin\theta_j\,\cos\phi_j,\sin\theta_j\,\sin\phi_j,\cos\theta_j)$
determines the direction of the particle magnetic moment, with 
$\theta_j$ and $\phi_j$ the polar and azimuthal angles. 

Introducing now the lattice sums $T_{ij}^{\alpha\beta\gamma}$ and after 
some algebra one obtains for the three components of the dipole field 
\begin{eqnarray}
\fl B_i^{x,\mathrm{dip}}=
\mu_{at}\,\sum_j N_j \Bigg\{ \Big(2\,T_{ij}^{520}-T_{ij}^{502}
-z_{ij0}^2\,T_{ij}^{500}\Big)\,m_j^x+3\,T_{ij}^{511}\,m_j^y
+3\,z_{ij0}\,T_{ij}^{510}\,m_j^z \nonumber \\ \fl 
+\frac{5}{2} \bigg[\Big(14\,T_{ij}^{940}-10\,T_{ij}^{720}-7\,
T_{ij}^{922}+T_{ij}^{702}+z_{ij0}^2\,(T_{ij}^{700}-7\,T_{ij}^{920})
+\frac{4}{5}\,T_{ij}^{500}\Big)\,m_j^x \nonumber \\ \fl 
+(21\,T_{ij}^{931}-9\,T_{ij}^{711})\,m_j^y+z_{ij0}\,(21\,T_{ij}^{930}
-9\,T_{ij}^{710})\,m_j^z \bigg]\,\Big(\la x_i^2\ra +\la x_j^2\ra \Big)
\nonumber \\ \fl 
+\frac{5}{2} \bigg[\Big(14\,T_{ij}^{922}-2\,T_{ij}^{720}-7\,
T_{ij}^{904}+5\,T_{ij}^{702}+z_{ij0}^2\,(T_{ij}^{700}-7\,T_{ij}^{902})
-\frac{2}{5}\,T_{ij}^{500}\Big)\,m_j^x \nonumber \\ \fl 
+(21\,T_{ij}^{913}-9\,T_{ij}^{711})\,m_j^y+z_{ij0}\,(21\,T_{ij}^{912}
-3\,T_{ij}^{710})m_j^z  \bigg]\,\Big(\la y_i^2\ra +\la y_j^2\ra \Big)
\nonumber \\ \fl 
+\frac{5}{2} \bigg[\Big(14\,z_{ij0}^2\,T_{ij}^{920}-2\,T_{ij}^{720}
-7\,z_{ij0}^2\,T_{ij}^{902}+T_{ij}^{702}+z_{ij0}^2\,(5\,T_{ij}^{700}
-7\,z_{ij0}^2\,T_{ij}^{900})-\frac{2}{5}\,T_{ij}^{500}\Big)\,m_j^x 
\nonumber \\ \fl 
+(21\,z_{ij0}^2\,T_{ij}^{911}-3\,T_{ij}^{711})\,m_j^y+z_{ij0}\,(21\,
z_{ij0}^2\,T_{ij}^{910}-9\,T_{ij}^{710})\,m_j^z\bigg]\,\Big(
\la z_i^2\ra +\la z_j^2\ra \Big)\Bigg\} \;, \label{e24} \\ 
\fl B_i^{y,\mathrm{dip}}=\mu_{at}\,\sum_j N_j \Bigg\{ 3\,T_{ij}^{511}\,m_j^x+
\Big(2\,T_{ij}^{502}-T_{ij}^{520}-z_{ij0}^2\,T_{ij}^{500}\Big)\,m_j^y
+3\,z_{ij0}\,T_{ij}^{501}\,m_j^z \nonumber \\ \fl 
+\frac{5}{2} \bigg[\Big(14\,T_{ij}^{922}-2\,T_{ij}^{702}-7\,
T_{ij}^{940}+5\,T_{ij}^{720}+z_{ij0}^2\,(T_{ij}^{700}-7\,T_{ij}^{920})
-\frac{2}{5}\,T_{ij}^{500}\Big)\,m_j^y \nonumber \\ \fl 
+(21\,T_{ij}^{931}-9\,T_{ij}^{711})\,m_j^x+z_{ij0}\,(21\,
T_{ij}^{921}-9\,T_{ij}^{701})\,m_j^z  \bigg]\,
\Big(\la x_i^2\ra +\la x_j^2\ra \Big) \nonumber \\ \fl 
+\frac{5}{2} \bigg[\Big(14\,T_{ij}^{904}-10\,T_{ij}^{702}-7\,
T_{ij}^{922}+T_{ij}^{720}+z_{ij0}^2\,(T_{ij}^{700}-7\,T_{ij}^{902})
+\frac{4}{5}\,T_{ij}^{500}\Big)\,m_j^y \nonumber \\ \fl 
+(21\,T_{ij}^{913}-9\,T_{ij}^{711})\,m_j^x+z_{ij0}\,(21\,T_{ij}^{903}
-9\,T_{ij}^{701})\,m_j^z  \bigg]\,\Big(\la y_i^2\ra +\la y_j^2\ra \Big)
\nonumber \\ \fl 
+\frac{5}{2} \bigg[\Big(14\,z_{ij0}^2\,T_{ij}^{902}-2\,T_{ij}^{702}
-7\,z_{ij0}^2\,T_{ij}^{920}+T_{ij}^{720}+z_{ij0}^2\,(5\,T_{ij}^{700}
-7\,z_{ij0}^2\,T_{ij}^{900})-\frac{2}{5}\,T_{ij}^{500}\Big)\,m_j^y 
\nonumber \\ \fl 
+(21\,z_{ij0}^2\,T_{ij}^{911}-3\,T_{ij}^{711})\,m_j^x+z_{ij0}\,
(21\,z_{ij0}^2\,T_{ij}^{901}-9\,T_{ij}^{701})\,m_j^z\bigg]\,\Big
(\la z_i^2\ra +\la z_j^2\ra \Big)\Bigg\} \;, \label{e25} \\
\fl B_i^{z,\mathrm{dip}}=\mu_{at}\,\sum_j N_j \Bigg\{ 
\Big(2\,z_{ij0}^2\,T_{ij}^{500}-T_{ij}^{520}-T_{ij}^{502}\Big)\,m_j^z
+3\,z_{ij0}\,\Big(T_{ij}^{510}\,m_j^x+T_{ij}^{501}\,m_j^y\Big)
\nonumber \\ \fl +\frac{5}{2} \bigg[
\Big(14\,z_{ij0}^2\,T_{ij}^{920}-2\,z_{ij0}^2\,T_{ij}^{700}
-7\,T_{ij}^{940}+5\,T_{ij}^{720}-7\,T_{ij}^{922}+T_{ij}^{702}
-\frac{2}{5}\,T_{ij}^{500}\Big)\,m_j^z \nonumber \\ \fl 
+z_{ij0}\,(21\,T_{ij}^{930}-9\,T_{ij}^{710})\,m_j^x
+z_{ij0}\,(21\,T_{ij}^{921}-3\,T_{ij}^{701})\,m_j^y \bigg]\,
\Big(\la x_i^2\ra +\la x_j^2\ra \Big) \nonumber \\ \fl 
+\frac{5}{2} \bigg[\Big(14\,z_{ij0}^2\,T_{ij}^{902}-2\,z_{ij0}^2\,
T_{ij}^{700}-7\,T_{ij}^{904}+5\,T_{ij}^{702}-7\,T_{ij}^{922}+T_{ij}^{720}
-\frac{2}{5}\,T_{ij}^{500}\Big)\,m_j^z \nonumber \\ \fl 
+z_{ij0}\,(21\,T_{ij}^{912}-3\,T_{ij}^{710})\,m_j^x+z_{ij0}\,(21\,
T_{ij}^{903}-9\,T_{ij}^{701})\,m_j^y \bigg]\,
\Big(\la y_i^2\ra +\la y_j^2\ra \Big)\nonumber \\ \fl 
+\frac{5}{2} \bigg[\Big(z_{ij0}^2\,\Big(14\,z_{ij0}^2\,T_{ij}^{900}
-10\,T_{ij}^{700}-7\,T_{ij}^{920}-7\,T_{ij}^{902}\Big)+T_{ij}^{720}
+T_{ij}^{702}+\frac{4}{5}\,T_{ij}^{500}\Big)\,m_j^z \nonumber \\ \fl 
+z_{ij0}\,(21\,z_{ij0}^2\,T_{ij}^{910}-9\,T_{ij}^{710})\,m_j^x
+z_{ij0}\,(21\,z_{ij0}^2\,T_{ij}^{901}-9\,T_{ij}^{701})\,m_j^y
\bigg]\,\Big(\la z_i^2\ra +\la z_j^2\ra \Big)\Bigg\} \;. \label{e26}
\end{eqnarray} 

The lateral extensions $L_x$ and $L_y$ of the unit cell and the 
components $x_{ij0}$ and $y_{ij0}$ of the interparticle distances 
within a unit cell 
are involved in a complicated manner in the lattice sums
$T_{ij}^{\alpha\beta\gamma}$, whereas the dependence on 
the vertical distances $z_{ij0}$
appears explicitly. Note that the contributions from mirror
particles $i=j$ located in different unit cells are taken into account.
The expressions for the dipole field components, 
equations~\ref{e24} -- \ref{e26}, are simplified appreciably if all 
particles are located in the same plane ($z_{ij0}=0$). In this case 
$B_i^{z,\mathrm{dip}}\propto m_j^z$, hence 
the $z$ component of the dipole field vanishes for an in-plane
magnetization ($m_j^z=0$). As before, the point-dipole sum is recovered by 
setting $\la x_i^2\ra = 0$, etc.
The dipole energy $E_\mathrm{dip}$ per unit cell is obtained 
from equation~\ref{e1} by performing the sum over all particles
$i=1\ldots n$ in the unit cell.

\end{document}